\documentclass[twocolumn,showpacs,preprintnumbers,amsmath,amssymb,superscriptaddress]{revtex4-1}

\usepackage{graphicx}
\usepackage{dcolumn}
\usepackage{bm}
\usepackage{tabularx}
\usepackage{makecell}
\usepackage{braket}
\usepackage{lipsum}
\usepackage{ulem}
\usepackage{amsmath,amssymb} 
\newcolumntype{M}{>{\centering\arraybackslash}m{1.85cm}}
\usepackage[export]{adjustbox}
\usepackage{longtable}
\usepackage{float}
\usepackage{xcolor}
\usepackage{color}   
\usepackage{hyperref}
\hypersetup{
	colorlinks=true, 
	linkcolor=blue,  
}

\makeatletter
\newcommand{\colorcaption}[2][]{%
	\begingroup%
	\renewcommand{\@caption@fignum@sep}{ (Color online). }%
	\caption[#1]{#2}%
	\endgroup%
}
\makeatother

\begin{document}
	\title{Systematic shell-model study of $^{99-129}$Cd isotopes and isomers in neutron-rich $^{127-131}$In isotopes}
	
	\author{ Deepak Patel}
	\address{Department of Physics, Indian Institute of Technology Roorkee, Roorkee 247667, India}
 	\author{Praveen C. Srivastava}
  \email{Corresponding author: praveen.srivastava@ph.iitr.ac.in}
	\address{Department of Physics, Indian Institute of Technology Roorkee, Roorkee 247667, India}
	\author{Noritaka Shimizu}
	\address{Center for Computational Sciences, University of Tsukuba, 1-1-1, Tennodai Tsukuba,\\ Ibaraki 305-8577, Japan}
	\author{Yutaka Utsuno}
	\address{Advanced Science Research Center, Japan Atomic Energy Agency, Tokai, Ibaraki 319-1195, Japan}
    \address{Center for Nuclear Study, University of Tokyo, 7-3-1 Hongo, Bunkyo-ku, \\Tokyo 113-0033, Japan}

	\date{\hfill \today}
	
	
	\begin{abstract}
		Systematic shell-model calculations are presented for odd-mass Cd isotopes with $N=51-81$
  utilizing a combination of a $G$-matrix interaction and a semiempirical one. 
        The excited energy spectra and electromagnetic transition probabilities are compared with the recently available experimental data. 
We have found that the observed quadrupole moments in the $11/2^-_1$ states that linearly change with the neutron number are well accounted for by the dominance of prolate shapes throughout the Cd isotope chain. 
    We have also described the properties of several isomeric states in neutron-rich $^{127-131}$In isotopes 
that were recently observed in Jyv\"askyl\"a.
	\end{abstract}

	
	\maketitle
	
	\section{Introduction}
	Recently, the $Z=50$ mass region emerged as an interesting case for nuclear structure studies \cite{Yordanov1,Togashi,Nara_Singh,Vaman,Piet}. The accessibility of new experimental techniques provides a better environment to perform new experiments in this region. There are three longest isotopic chains near $^{100}$Sn, i.e., cadmium ($Z=48$), tin ($Z=50$), and tellurium ($Z=52$), isotopes where recent experiments have been performed to study spectroscopic properties of these nuclei. One of the interests in the structure of these isotopes is the role of the additional valence proton particles or holes, and collective properties in even-$A$ Cd isotopes are discussed in Refs. \cite{Nomura,Zuker,Siciliano}, for instance.

    Odd-$A$ isotopes provide more detailed information on nuclear structure. With regard to Cd isotopes, Park $et$ $al.$ \cite{Park} studied neutron-deficient nuclei $^{99}$Cd and assigned new excited states such as $(1/2^+)$ and $(3/2^+)$ at 883 and 1077 keV energies, respectively. Yordanov $et$ $al.$ \cite{Yordanov1} studied the neutron-deficient Cd isotopes using high-resolution laser spectroscopy. In another work, their group investigated the $^{107-129}$Cd isotopes \cite{Yordanov}. They have mainly focused on studying the behavior of quadrupole and magnetic moments of the $11/2^-$ states. In the medium mass region,  Kisyov $et$ $al.$ \cite{Kisyov} performed fast-timing measurements in $^{103,105,107}$Cd isotopes. They have also interpreted that the $7/2^+_1$ state arises from a single-particle excitation. Ashley $et$ $al.$ \cite{Ashley} deduced the mean-lifetime of yrast $15/2^-$ state (at 1360 keV) in $^{107}$Cd. Stuchbery $et$ $al.$ \cite{Stuchbery} measured the $g$ factors of low-lying states in $^{111}$Cd and $^{113}$Cd and showed the sensitivity of this observable for the collective nature of these nuclei. Besides this, Coombes $et$ $al.$ \cite{Coombes} studied the structure of $^{111}$Cd through angular correlation, lifetime, and $g$-factor measurements. In the neutron-rich region, Hoteling $et$ $al.$ \cite{Hoteling} identified the microsecond isomers in $^{125-128}$Cd isotopes. In this region, Manea $et$ $al.$ \cite{Manea} measured the masses of neutron-rich Cd isotopes and isomers near $N=82$. Using the phase-imaging ion cyclotron resonance technique, they have also established the inversion of $11/2^-$ and $3/2^+$ states in $^{129}$Cd.

	Several experiments have also been performed for the neutron-rich indium ($Z=49$) isotopes in recent years. The indium isotopes from $^{127}$In to $^{131}$In are particularly interesting to probe shell structure around $N=82$. Nesterenko $et$ $al.$ \cite{Nesterenko} studied the isomeric states in $^{128}$In using the JYFLTRAP Penning trap at the IGISOL facility. They have also discovered a new isomer $(16^+)$ in $^{128}$In at 1797.6(20) keV. In Ref. \cite{Lorenz}, a spectroscopic study of $\beta$ decay of $^{127}$Cd and excited states of $^{127}$In was conducted. They have also compared the experimental excited states of $^{127}$In with the shell-model results using the jj45 interaction. Recently, Vernon $et$ $al.$ \cite{Vernon1} performed precision laser spectroscopy measurements in indium isotopes and observed an abrupt change in the magnetic moment of the $9/2^+$ ground state (g.s.) at the neutron magic number $N=82$.

    It is of great importance how those observables are described and how the excited states that have yet to be observed are predicted in microscopic theories. While the shell-model calculation is a promising choice, its application to the mid-shell Cd isotopes was challenging because of the huge dimension of the shell-model Hamiltonian matrix compared to that of the Sn isotopes, for which our group has recently performed shell-model calculations in the 50-82 valence shell and described different seniority states of $^{119-126}$Sn isotopes \cite{Srivastava}. At present, advanced computing resources provide an opportunity to carry out such demanding calculations.

    In the present study, we have performed large-scale shell-model calculations for odd-$A$ Cd isotopes. Besides energy levels and electromagnetic transition probabilities, of particular interest are the quadrupole moments of the $11/2^-_1$ states that follow a linear increase with the neutron number \cite{Yordanov}. Since this behavior is well described by the seniority scheme, the data have been regarded as a signature of the extreme shell-model states in a single-$j$ shell. In this work, we activate multiple valence orbitals for both protons and neutrons in large-scale shell-model calculations, and can examine whether the deformations associated with the large model space change the story. The structure of isomeric states in $^{127-131}$In are also studied in terms of the seniority ($v$).

	This paper is divided into the following sections. In Sec. \ref{section2}, the details about the interactions used in our calculations are briefly described. In Sec. \ref{section3}, we present our calculated results of energy spectra, electromagnetic transition probabilities, deformation of the $11/2_1^{-}$ band, and electromagnetic moments in the Cd chain. We also discuss the properties of isomeric states in neutron-rich In isotopes. Finally, we summarize our results and conclude the paper in Sec. \ref{section4}.
	
	\section{Theoretical Framework} \label{section2}

	In the present study, we performed the shell-model calculations of the odd-mass $^{99-129}$Cd and $^{127-131}$In isotopes using an interaction comprising a $G$-matrix-based one and a semiempirical one. The model space of the present study consists of two proton orbitals ($1p_{1/2}$ and $0g_{9/2}$) and five neutron orbitals ($1d_{5/2}$, $2s_{1/2}$, $1d_{3/2}$, $0g_{7/2}$, and $0h_{11/2}$) with the $^{88}$Sr inert core. In Ref.~\cite{Boelaert} the $G$-matrix interaction for this model space was constructed with the phenomenological corrections for neutron-deficient even-mass ($50\leq N \leq 58$) Cd isotopes, and their collective features were discussed. Using this interaction, some of the present authors discussed the low-lying states and the isomeric $8^+$ states of even-mass Cd isotopes in Ref.~\cite{Patel_NPA_evenCd}. However, we found that this $G$-matrix interaction underestimates the $2^+_1$ excitation energies of Cd isotopes around $A=120$ because this interaction was only fitted in the neutron-deficient region. In order to solve this problem, we replace the neutron-neutron part of this interaction with the SNBG1 interaction, which was proposed in \cite{SNBG1}. The SNBG1 interaction was constructed in the $50< N<82$ model space with the $^{100}$Sn inert core. It was based on the $G$ matrix and was $\chi^2$ fitted for the 133 experimental data of Sn isotopes \cite{SNBG1}. Therefore, the combined interaction reproduces the same result of the Sn isotopes as that of the SNBG1 interaction, which shows excellent agreement with the experimental data.

\begin{figure}[h]
		\includegraphics[width=82mm]{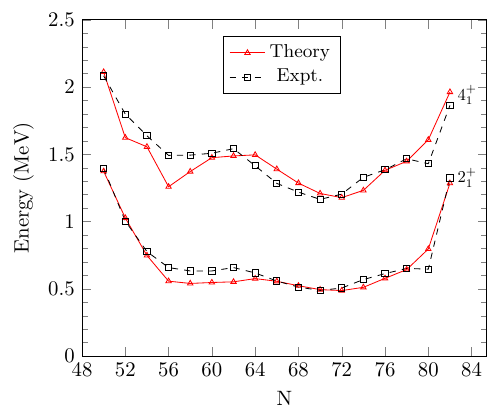}
  \includegraphics[width=82mm]{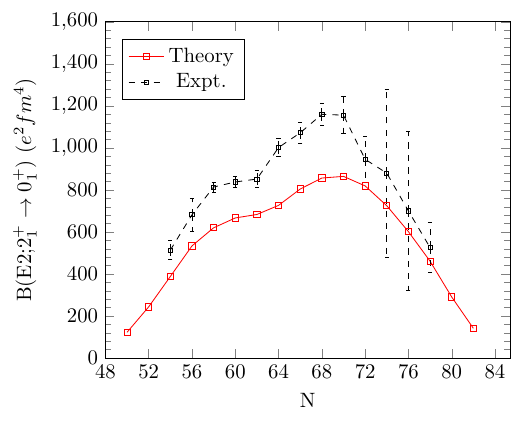}
		\caption{\label{fig_even} Shell-model and experimental \cite{NNDC} energy levels of $2_1^+$ and $4_1^+$ states (top panel), and $B(E2: 2_1^+ \rightarrow 0_1^+)$ values  for even-mass $^{98-130}$Cd (bottom panel). The experimental data for $B(E2)$ values are taken from 	\cite{Pritychenko}.}
 \end{figure}

    Figure \ref{fig_even} shows the $2^+_1$ and $4^+_1$ energies and the $B(E2; 2^+_1 \to 0^+_1)$ values of even-mass Cd isotopes. Throughout this study, we use the $E2$ effective charges of $(e_{\pi}, e_{\nu})=(1.6e, 0.8e)$, following the values employed in Ref. \cite{SNBG_ef}. The calculated $B(E2)$ values follow the experimental trend well, including the peak at around $N=70$, although they are somewhat smaller than the data on the whole. The good agreement with the experimental results proves the validity of the current shell-model study.

 The shell-model code KSHELL \cite{KShell} was used to diagonalize the shell-model Hamiltonian matrices for our calculations. It was performed without any truncation in the model space. The largest $M$-scheme dimension is 0.9 $\times$ 10$^9$ for $^{113,115}$Cd.

	\section{Results and Discussion} \label{section3}

\begin{figure*}
		\includegraphics[width=182mm]{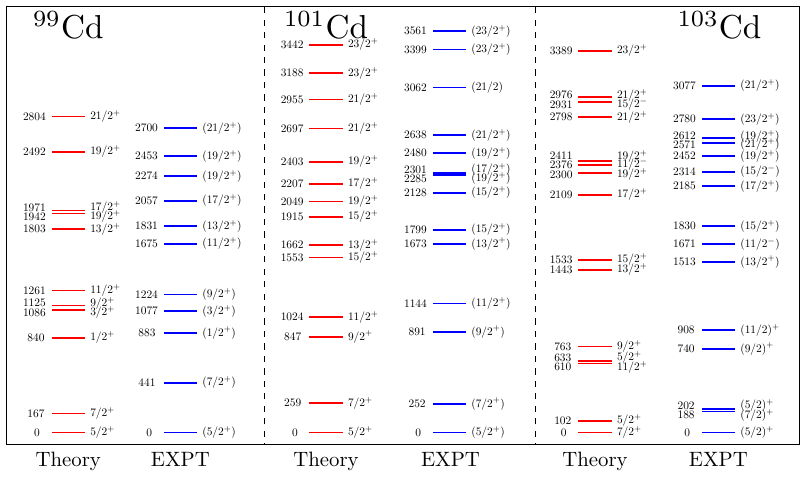}
		\includegraphics[width=182mm]{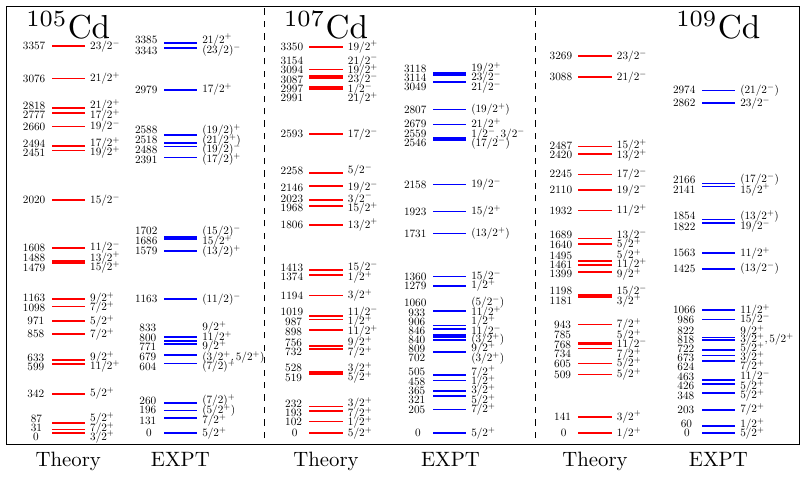}
		\caption{\label{fig1} Comparison between calculated and experimental \cite{NNDC} energy levels for odd-mass $^{99-109}$Cd.}
	\end{figure*}

\begin{figure*}
		\includegraphics[width=182mm]{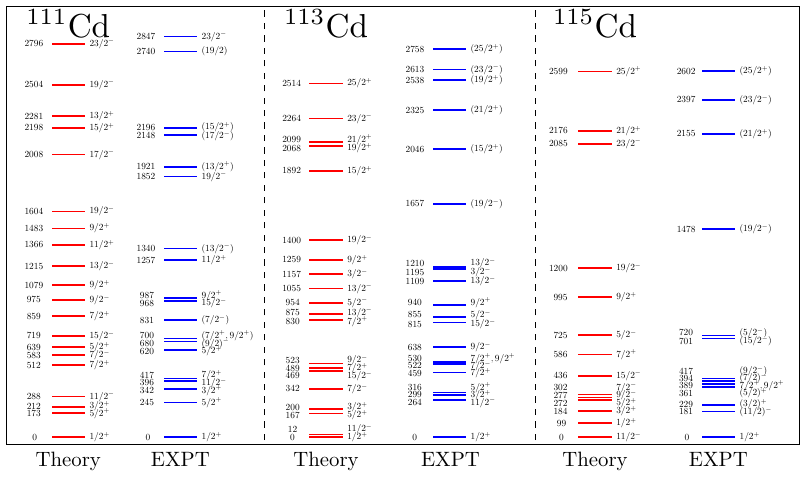}
		\includegraphics[width=182mm]{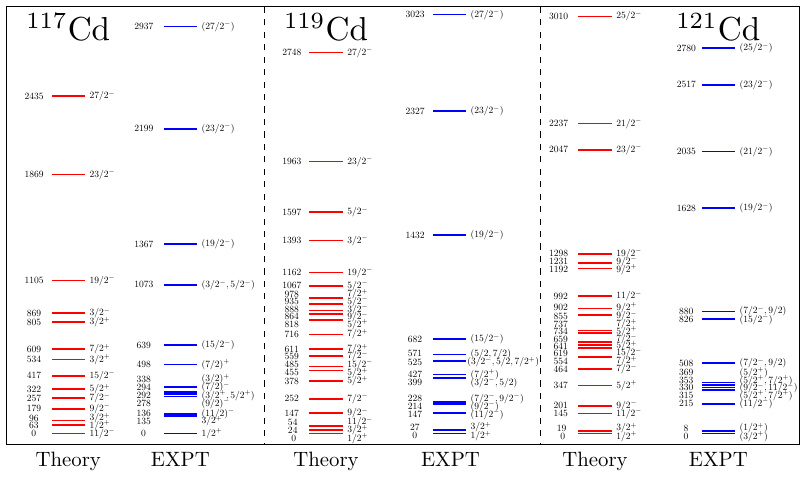}
		\caption{\label{fig2} Comparison between calculated and experimental \cite{NNDC} energy levels for odd-mass $^{111-121}$Cd isotopes.}
	\end{figure*}

	In this section, we present shell-model results for odd-mass $^{99-129}$Cd isotopes and compare them with the experimental data. Figures \ref{fig1}-\ref{fig4}, show the low-lying shell-model energy spectra of $^{99-129}$Cd isotopes compared to the experimental data. The electric quadrupole and magnetic dipole transition probabilities are reported in Tables \ref{BE2} and \ref{B(M1)}, respectively.

	\subsection{Excitation spectra and Electromagnetic transition probabilities}
	
	Here, we discuss the energy spectra and electromagnetic transition probabilities in the odd-mass Cd isotopic chain of neutron numbers $N=51-81$. The experimental data are compared with our shell-model results.
	
	{\bf $^{99}$Cd}: The shell-model excitation energies of the  $9/2^+_1$, $13/2^+_1$, $17/2^+_1$, and $19/2^+_2$ states are 
  in good agreement with the experimental data. We have compared the location of two other states $(1/2^+)$ and $(3/2^+)$ with the experimental excitation energies 883 and 1077 keV as reported in Ref. \cite{Park}. These states are reproduced well by the shell-model result showing the predominant $\pi(g_{9/2}^{-2})\otimes\nu(d_{5/2}^1)$ character. The shell-model $5/2^+_1$, $9/2^+_1$, $13/2^+_1$, $17/2^+_1$, $19/2^+_2$ and $21/2^+_1$ states are dominated by the $\pi(g_{9/2}^{-2})\otimes\nu(d_{5/2}^1)$ configuration. For the isomeric state $17/2^+_1$, the calculated $B(E2; 17/2^+_1\rightarrow13/2^+_1)$ value is close to the experimental data as reported in Table \ref{BE2}. On the other hand, the $7/2^+_1$ and $11/2^+_1$ states show $\pi(g_{9/2}^{-2})\otimes\nu(g_{7/2}^1)$ character. These results support the previous calculations, as reported in Ref. \cite{Lipoglavsek}.

  {\bf $^{101}$Cd}: The $^{101}$Cd has two proton holes and three neutron particles on the doubly magic  $^{100}$Sn core. The measured energy levels are reasonably reproduced by the shell-model calculation. The $5/2^+_1$, $9/2^+_1$, and $13/2^+_1$ states are dominated by the $0^+$, $2^+$, and $4^+$ core coupled to a neutron $d_{5/2}$ state, respectively. Also, the $7/2^+_1$, $11/2^+_1$, $15/2^+_1$, and $19/2^+_1$ states (members of the same band) are dominated by the $0^+$, $2^+$, $4^+$, and $6^+$ core coupled to an unpaired neutron in $g_{7/2}$, respectively. Alber \textit{et al.} \cite{Alber} identified the $19/2^+_1$ state as an isomer that decays to the $15/2^+_1$ state with a half-life 4.6(4) ns, which corresponds to a small $B(E2)$ value of 0.160(20) W.u. From the shell model calculations, both the initial ($19/2^+$) and final ($15/2^+$) states are favored by seniority $v=3$, hence because of the same seniority their $E2$ transition is hindered. As reported in Table \ref{BE2}, shell-model calculations also predict a relatively small $B(E2)$ value, which supports the $19/2^+$ state as a seniority isomer.

  {\bf $^{103}$Cd}: In the Cd chain, $^{103}$Cd lies between the transitional region of nearly spherical nuclei $^{98}$Cd and deformed nuclei with more valence neutrons. For $^{103}$Cd, the g.s. $(5/2)^+$ is not reproduced by the shell model; this state lies 102 keV higher than the proposed g.s. $7/2^+$ with our calculation. Experimentally, the $11/2^-$ state has been observed from $^{103}$Cd onward in the Cd chain. However, our calculated $11/2^-$ state lies at higher energy than the experimental data. Similar to $^{101}$Cd, the $(19/2^+)$ state is known to be an isomer with a small $B(E2)$ value of 0.162(25) W.u. Our calculation fails in reproducing such a small $B(E2)$ value. This could be  due to the fact that the configurations of $15/2^+$ [$\pi(g_{9/2}^{-2})\otimes\nu(d_{5/2}^4g_{7/2}^1)$] and $19/2^+$ [$\pi(g_{9/2}^{-2})\otimes\nu(d_{5/2}^3g_{7/2}^2)$] states are the not same as in case of $^{101}$Cd.

 {\bf $^{105}$Cd}: The $5/2^+$ state is assigned as the ground state in $^{105}$Cd \cite{Laulainen,Chapman},
 whereas it lies 87 keV higher than the proposed g.s. $3/2^+_1$ state by the shell-model calculation. The excited $(21/2^+)$ state is established as an isomeric state with $T_{1/2}=4.5(5)$ $\mu$s \cite{Frenne}. This $21/2^+_1$ isomer [$\pi(g_{9/2}^{-2})\otimes\nu(d_{5/2}^5g_{7/2}^2)$] decays via $E2$ transition into the $17/2^+_1$ state. As reported in Table \ref{BE2}, the shell-model calculated $B(E2;21/2^+_1\rightarrow17/2^+_1)$ is close to the experimental data, with $21/2^+_1$ 
 dominated by an unpaired neutron in $d_{5/2}$ coupled to the fully aligned $8^+$ state for protons. 
	
	{\bf $^{107}$Cd}: 
 The experimental $5/2^+$ ground state is reproduced by the shell-model calculation. The calculated $7/2^+_1$ state 
 shows a good agreement with the experimental data. With our calculations, this state is formed by the dominance of an unpaired neutron in $\nu(g_{7/2})$, and it decays via $E2+M1$ transition to the g.s. $5/2^+_1$; the corresponding $B(E2)$ and $B(M1)$ values are reported in Tables \ref{BE2} and \ref{B(M1)}, respectively. The location of calculated $9/2^+_1$, $1/2^+_2$, $11/2^+_1$ $1/2^+_3$, $15/2^-_1$, $13/2^+_1$, $15/2^+_1$, $19/2^-_1$, and $17/2^-_1$ states are also comparable with the experimental data. The $9/2^+_1$ state decays via $E2$ transitions into $5/2^+_1$ and $7/2^+_1$ states. As reported in Table \ref{BE2}, the calculated $B(E2)$ values show reasonable agreement with the experimental data corresponding to decay from the $9/2^+_1$ state.

{\bf $^{109}$Cd}: The $5/2^+$ ground state is not reproduced by our calculation. However, the main configuration of this state comes from the odd neutron in $\nu(d_{5/2})$. We have also reported the $B(E2)$ transitions for high spin states $19/2^-_1$ and $23/2^-_1$ in Table \ref{BE2}. The shell-model predicted $B(E2;19/2^-_1\rightarrow15/2^-_1)$ value is almost half of the experimental value, although the experimental uncertainty is rather large.

	\begin{figure*}
		\includegraphics[width=182mm]{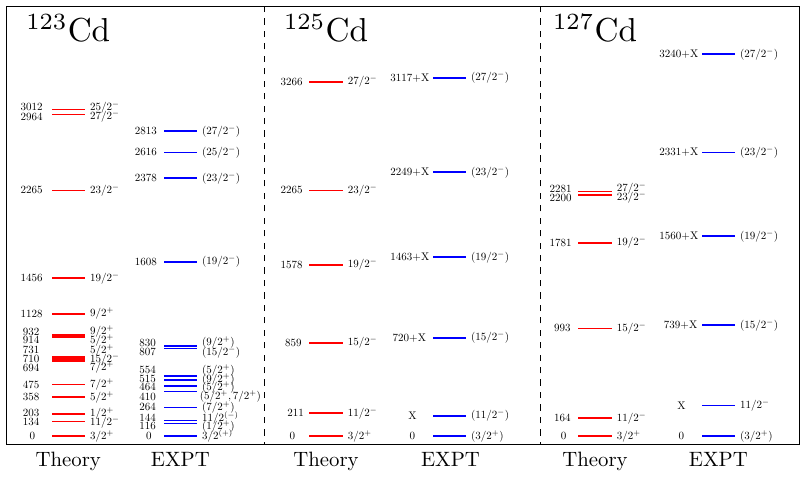}
		\includegraphics[width=70mm]{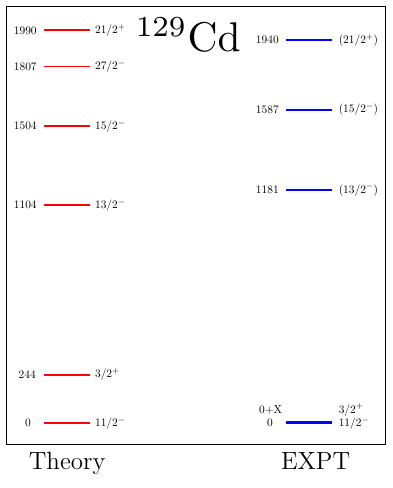}
		\caption{\label{fig4} Comparison between calculated and experimental \cite{NNDC} energy levels for odd-mass $^{123-129}$Cd isotopes.  In the case of $^{125,127}$Cd, we assume $X=186$ and 283.3 keV, respectively, and $X=0$ keV for $^{129}$Cd isotopes.}
	\end{figure*}
	
	\begin{table}
		\centering
		\caption{Comparison of theoretical and experimental $B(E2)$ transitions (in $e^2\textrm{fm}^4$) \cite{NNDC_NUDAT} in $^{99-129}$Cd isotopes. The effective charges are taken as $(e_\pi, e_\nu)=(1.6, 0.8)e$ \cite{SNBG_ef}.
  }
		\begin{ruledtabular}
		\begin{tabular}{cccccc}
			
		Isotope	&	$J_i^{\pi} \rightarrow J_f^{\pi}$   & Theory  &	Expt. 	 \\  
			\hline
			$^{99}$Cd	&	$7/2^{+}_{1}$$\rightarrow$$5/2^{+}_{1}$   & 9.8 & - \\ 
			&	$1/2^{+}_{1}$$\rightarrow$$5/2^{+}_{1}$   & 350.0 & - \\ 
			&	$17/2^{+}_{1}$$\rightarrow$$13/2^{+}_{1}$    & 65.8 &  68(3)  \\ 
			$^{101}$Cd	&   $7/2^{+}_{1}$$\rightarrow$$5/2^{+}_{1}$   & 23.3 & - \\ 
			&	$9/2^{+}_{1}$$\rightarrow$$5/2^{+}_{1}$    & 290.8 & - \\ 
			&	$9/2^{+}_{1}$$\rightarrow$$7/2^{+}_{1}$     & 41.8 & - \\ 
			&   $19/2^{+}_{1}$$\rightarrow$$15/2^{+}_{1}$   & 27.9 &  4.5(6)  \\ 
			$^{103}$Cd	&	$7/2^{+}_{1}$$\rightarrow$$5/2^{+}_{1}$    & 2.6 & - \\ 
			&	$9/2^{+}_{1}$$\rightarrow$$5/2^{+}_{1}$     & 489.7 & - \\ 
			&	$9/2^{+}_{1}$$\rightarrow$$7/2^{+}_{1}$      & 19.0 & - \\ 
			&	$19/2^{+}_{1}$$\rightarrow$$15/2^{+}_{1}$     &  123.4 &  4.65(72)  \\ 
			$^{105}$Cd  &	$11/2^{+}_{1}$$\rightarrow$$7/2^{+}_{1}$     & 735.9 & - \\ 
			&	$13/2^{+}_{1}$$\rightarrow$$11/2^{+}_{1}$     & 135.1 & - \\ 
			&	$15/2^{+}_{1}$$\rightarrow$$11/2^{+}_{1}$     & 812.7 & - \\ 
			&	$21/2^{+}_{1}$$\rightarrow$$17/2^{+}_{1}$     & 2.9 & 2.35(26) \\ 
			$^{107}$Cd	&	$7/2^{+}_{1}$$\rightarrow$$5/2^{+}_{1}$    & 797.4 & 120.7(91) \\ 
			&	$9/2^{+}_{1}$$\rightarrow$$5/2^{+}_{1}$    & 326.5 & 331.9(1810) \\ 
			&	$9/2^{+}_{1}$$\rightarrow$$7/2^{+}_{1}$     & 665.6 & 482.8(2716) \\ 
			&	$11/2^{+}_{1}$$\rightarrow$$7/2^{+}_{1}$     & 640.2 & 573.3(1810) \\ 
			&	$15/2^{-}_{1}$$\rightarrow$$11/2^{-}_{1}$    & 892.9 & 718.1(392) \\ 
			&	$15/2^{+}_{1}$$\rightarrow$$11/2^{+}_{1}$    & 841.6 & 663.8(1810) \\ 
			&	$19/2^{-}_{1}$$\rightarrow$$15/2^{-}_{1}$    & 926.1 & 754.3(2414) \\ 
			$^{109}$Cd	&	$1/2^{+}_{1}$$\rightarrow$$5/2^{+}_{1}$    & 36.8 & 6.19(93) \\ 
   		    &	$15/2^{-}_{1}$$\rightarrow$$11/2^{-}_{1}$   & 917.8 &  1447.4(588) \\ 
			&	$19/2^{-}_{1}$$\rightarrow$$15/2^{-}_{1}$   & 1009.2 & 2350.5(5567) \\ 
			&	$23/2^{-}_{1}$$\rightarrow$$19/2^{-}_{1}$   & 942.9 & 371 $>$ \\ 
			$^{111}$Cd	&	$5/2^{+}_{1}$$\rightarrow$$1/2^{+}_{1}$   & 1.7 & 7.03(6) \\ 
			&	$3/2^{+}_{1}$$\rightarrow$$1/2^{+}_{1}$    & 530.4 & 370.7(444) \\ 
			$^{113}$Cd	&	$3/2^{+}_{1}$$\rightarrow$$1/2^{+}_{1}$    & 113.4 & 649.0(2596) \\ 
			&	$5/2^{+}_{1}$$\rightarrow$$1/2^{+}_{1}$    & 2.1 & 12.1(8) \\ 
			&	$7/2^{-}_{1}$$\rightarrow$$11/2^{-}_{1}$    & 1109.8 & 1434.3(714) \\ 
			$^{115}$Cd	&	$7/2^{-}_{1}$$\rightarrow$$11/2^{-}_{1}$   & 1160.9 & 1461.6(664) \\ 
			$^{117}$Cd	&	$7/2^{+}_{1}$$\rightarrow$$3/2^{+}_{1}$   & 13.0 & 15.0(34) \\ 
			&	$7/2^{-}_{1}$$\rightarrow$$11/2^{-}_{1}$   & 1057.8 & 1261.1(714) \\ 
			$^{119}$Cd	&	$7/2^{-}_{1}$$\rightarrow$$9/2^{-}_{1}$   & 1152.3 & - \\ 
			&	$9/2^{-}_{2}$$\rightarrow$$11/2^{-}_{1}$   & 7.9 & - \\ 
			$^{121}$Cd  &	$5/2^{+}_{1}$$\rightarrow$$3/2^{+}_{1}$    & 403.6 & - \\ 
			$^{123}$Cd	&	$7/2^{+}_{1}$$\rightarrow$$3/2^{+}_{1}$   & 214.3 & 3.6(+9-7) \\ 

			\label{BE2}
		\end{tabular}
		\end{ruledtabular}
	\end{table}

	\begin{table}
		\centering
		\caption{Comparison between theoretical and experimental $B(M1)$ (in $\mu_N^{2}$) values in odd-mass cadmium isotopes.  In the shell-model calculations, the spin part is quenched by a factor 0.7.
  The experimental data are taken from Ref. \cite{NNDC_NUDAT}.
  }
		\begin{ruledtabular}
		\begin{tabular}{cccccc}
			
			Isotope	&	$J_i^{\pi} \rightarrow J_f^{\pi}$  & Theory &	Expt.	\\ 
			\hline
			$^{105}$Cd	&	$7/2_{1}^{+}$$\rightarrow$$5/2_{1}^{+}$   &  0.026 & 0.0082(5) \\ 
			$^{107}$Cd	&	$7/2_{1}^{+}$$\rightarrow$$5/2_{1}^{+}$ & 0.038 & 0.0057(4) \\ 
			&	$9/2_{1}^{+}$$\rightarrow$$7/2_{1}^{+}$   & 0.045 & 0.0023(14) \\ 
			&	$21/2_{1}^{-}$$\rightarrow$$19/2_{1}^{-}$ & 0.013 & 0.052(23) \\ 
			$^{109}$Cd  &	$3/2_{1}^{+}$$\rightarrow$$1/2_{1}^{+}$  &  0.008  &   -   \\ 
			$^{111}$Cd	&	$3/2_{1}^{+}$$\rightarrow$$1/2_{1}^{+}$  & 0.008 & 0.010(2) \\ 
			&	$3/2_{1}^{+}$$\rightarrow$$5/2_{1}^{+}$  & 0.008 & 0.0367(4) \\ 
			$^{113}$Cd	&	$3/2_{1}^{+}$$\rightarrow$$1/2_{1}^{+}$  & 0.017 & 0.045(14) \\ 
			&	$5/2^{+}_{1}$$\rightarrow$$3/2^{+}_{1}$    & 0.017 & 0.015(2) \\ 
			$^{115}$Cd	&	$3/2_{1}^{+}$$\rightarrow$$1/2_{1}^{+}$  & 0.031 & - \\ 
			$^{117}$Cd	&	$3/2_{1}^{+}$$\rightarrow$$1/2_{1}^{+}$  & 0.039 & $>$0.013 \\ 
			$^{119}$Cd	&	$3/2_{1}^{+}$$\rightarrow$$1/2_{1}^{+}$  & 0.042 & 0.039(7) \\ 
			&	$9/2_{1}^{-}$$\rightarrow$$11/2_{1}^{-}$   & 0.026 & 0.036 $>$ \\ 
			
			
		\end{tabular}
		\end{ruledtabular}
		\label{B(M1)}
	\end{table}

	{\bf $^{111}$Cd}: The experimentally assigned g.s. $1/2^+$ is reproduced by the shell-model calculation. The first excited state $5/2^+$ (at 245 keV) is an $E2$ isomer \cite{Garg}, which decays into $1/2^+_{\text{g.s.}}$. The calculated $B(E2)$ value is reported in Table \ref{BE2}. The second excited state $3/2^+_1$ is strongly populated by Coulomb excitation \cite{McGowan}, and this state is dominated by the odd neutron in $\nu(d_{3/2})$ with our calculation. We have also compared the transition probabilities of $\gamma$-transitions from this state with the experimental data, which may give more information about the structure of this state. The $3/2^+_1$ state decays via $E2$ transitions into the $1/2^+_{\text{g.s.}}$ state. Our calculated $B(E2;3/2^+_1\rightarrow 1/2^+_1)$ value is almost 1.4 times higher than the experimental value. There are two $M1$ transitions also from $3/2^+_1$ to $1/2^+_{\text{g.s.}}$ and $5/2^+_1$ states. As reported in Table \ref{B(M1)}, the calculated $B(M1;3/2^+_1\rightarrow1/2^+_1)$ value is comparable with the experimental value.

	{\bf $^{113}$Cd}: For this isotope, the observed g.s. $1/2^+$ is reproduced by the shell-model calculation. As reported in Ref. \cite{Warr}, this state is a quasistable state; the main configuration of this state is dominated by an odd neutron in the $s_{1/2}$ orbital in our calculation. The isomeric state $5/2^+_1$ decays via $M1$ and $E2$ transitions in $3/2^+_1$ and $1/2^+_{g.s.}$ states \cite{NNDC}, respectively. The calculated $B(E2;7/2^-_1\rightarrow11/2^-_1)$ value is in a reasonable agreement with the experimental value.
	
    {\bf $^{115}$Cd}: The experimental $1/2^+$ ground state is not reproduced by the shell model, and it lies 99 keV above the theoretical $11/2^-_1$ ground state. We have compared the experimentally observed $B(E2)$ value for the transition between $7/2^-_1$ and $11/2^-_1$ states with our calculation. The calculated $B(E2;7/2^-_1\rightarrow 11/2^-_1)$ value shows a reasonable agreement with the experimental value as in the case of $^{113}$Cd.
	
	{\bf $^{117}$Cd}: The observed $1/2^+$ ground state is not reproduced in our calculation; the shell model predicts this state 63 keV higher than the proposed $11/2^-_1$ ground state. Experimentally, the first excited $3/2^+$ state (at 135 keV) decays via an $M1$ transition to the g.s. $1/2^+$; the observed $B(M1;3/2^+_1\rightarrow1/2^+_{g.s.})$ value is  greater than 0.013 $\mu_N^2$ \cite{NNDC_NUDAT}. Shell-model calculation predicts this $B(M1)$ value as 0.039 $\mu_N^2$, which is consistent with the experimental value. From our calculation, the $B(E2;7/2^-_1\rightarrow11/2^-_1)$ transition also shows reasonable agreement with the experimental value. As shown in Table \ref{BE2}, the $B(E2; 7/2^-_1 \to 11/2^-_1)$ values are stable for $^{113,115,117}$Cd both in theory and in experiment. Since the $7/2^-_1$ and $11/2^-_1$ states are dominated by the $0^+$ and $2^+$ cores coupled to a unpaired neutron in $h_{11/2}$, respectively, this may be due to the fact that the quadrupole deformation of the core is saturated in the middle of the $N=50$-$82$ shell. An $E2$ transition occurs between $7/2^+_1$ and $3/2^+_1$ states. As reported in Table \ref{BE2}, our calculated $B(E2;7/2^+_1\rightarrow3/2^+_1)$ value is comparable with the experimental value.

\begin{figure*}
	\includegraphics[width=89mm]{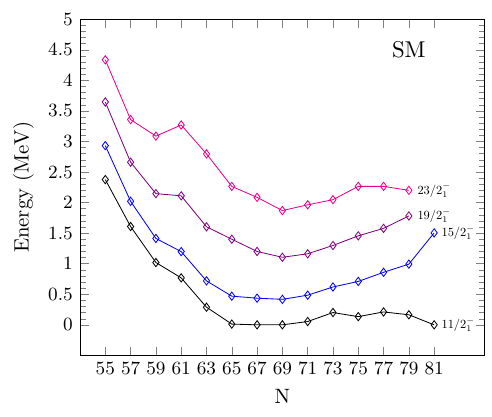}
	\includegraphics[width=89mm]{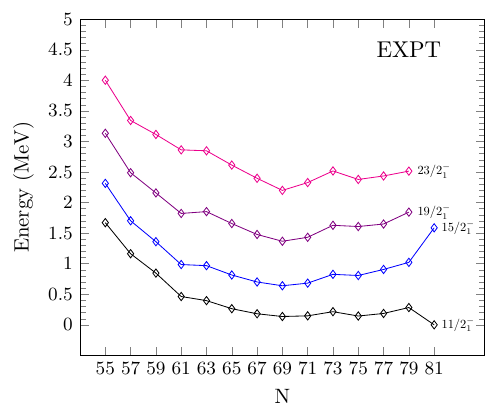}
	\caption{\label{band} Comparison between calculated (left panel) and experimental (right panel) \cite{NNDC} energies of the $11/2^-$ band in odd-mass $^{103-129}$Cd isotopes.}
\end{figure*}

 	\begin{figure}
		\includegraphics[width=82mm]{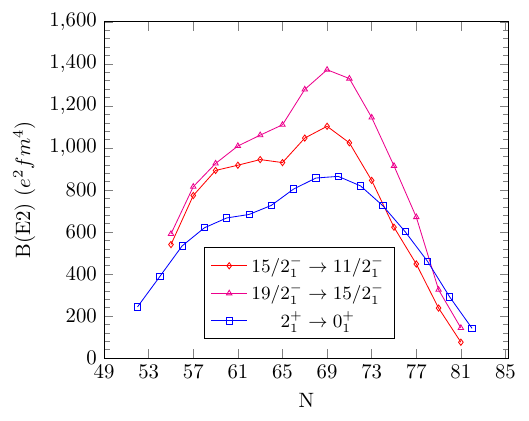}
		\caption{\label{be2_comparison} Shell-model $B(E2)$ transition probabilities for even- and odd-mass Cd isotopes.}
	\end{figure}

 \begin{figure*}
 \includegraphics[width=150mm]{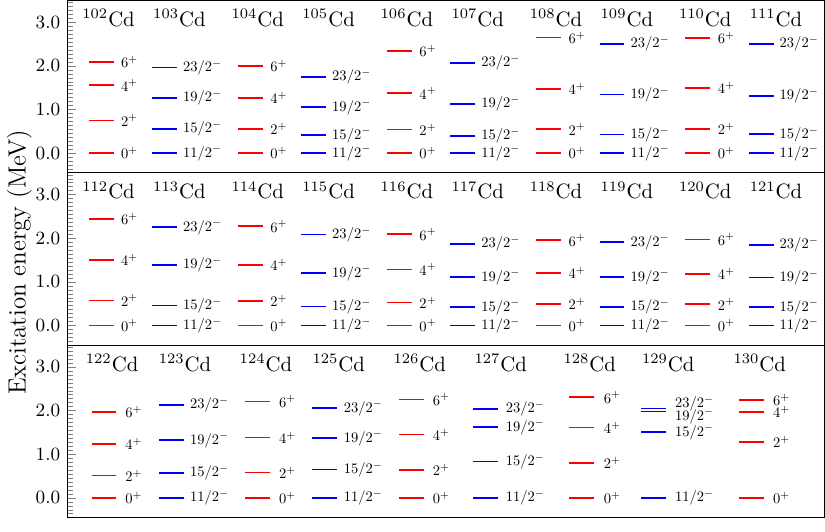}
	\caption{\label{band_comparision} Comparison of shell-model results of $11/2^-$ band in odd-mass Cd isotopes with the ground state band in even-mass Cd isotopes.}
\end{figure*}

{\bf $^{119}$Cd}: The ground-state spin $1/2^+$ is reproduced by the shell-model calculation. Our calculation reproduces also the first excited state $3/2^+$, which is dominated by an unpaired neutron in $\nu(d_{3/2})$. This $3/2^+$ state decays via $M1$ transition into the $1/2^+$ ground state. The corresponding $B(M1)$ value is comparable with the experimental data as reported in Table \ref{B(M1)}. The $9/2^-_1$ state and $\beta$-decaying isomer $11/2^-_1$ show reasonable agreement with the experimental data from our calculations. As reported in Ref. \cite{Garg}, the unconfirmed $(7/2^-,9/2^-)$ state at 228 keV is assigned as an $E2$ isomer that decays via $7/2^-\rightarrow9/2^-$ ($E_\gamma=14.3$ keV) or $9/2^-\rightarrow11/2^-$ ($E_\gamma=81.7$ keV) transitions, respectively. The shell-model $7/2^-_1$ state shows good agreement with the experimental data for this state. On the basis of these results, we predict the unconfirmed $(7/2^-,9/2^-)$ state could be $7/2^-_1$.

{\bf $^{121}$Cd}: The tentative g.s. $3/2^+$ lies 19 keV higher than the proposed g.s. $1/2^+$ with our calculation. This $3/2^+_1$ state ($\pi(g_{9/2}^{-2})\otimes\nu(g_{7/2}^6d_{3/2}^1h_{11/2}^8)$) is dominated by an odd neutron in the $\nu(d_{3/2})$ orbital, while the isomeric state $11/2^-_1$ ($\pi(g_{9/2}^{-2})\otimes\nu(g_{7/2}^6d_{3/2}^2h_{11/2}^7)$) \cite{Fogelberg,Fogelberg1} is formed by an unpaired neutron in the $\nu (h_{11/2})$ orbital. The spin of an $E2$ isomer $(5/2^+,7/2^+)$ at 314.5 keV is experimentally unconfirmed; it decays into g.s. $(3/2^+)$. The location of the shell-model calculated $5/2^+_1$ state is comparable with this state. 

{\bf $^{123}$Cd}: The tentative g.s. $3/2^{(+)}$ is reproduced with our calculation. The $\beta$-decaying $11/2^-$ 
 isomer is also in good agreement with the experimental data and our calculation. The locations of $1/2^+_1$ and $11/2^-_1$ states are reversed, and the $1/2^+_1$ state lies 87 keV higher than the experimental data with our calculations. The calculated $15/2^-_1$ state shows a reasonable agreement with the experimental data. 
The major contribution of this state comes from the coupling of the valence neutron in $\nu(h_{11/2})$ to the spin $I^{\pi}=2^+$  due to proton-hole pair breaking in $\pi(g_{9/2})$.

{\bf $^{125}$Cd}: The order of all excited states is reproduced with the shell-model calculation. The tentative g.s. $(3/2^+)$ is dominated by odd-neutron in $\nu(d_{3/2})$. For this isotope, the exact location of the excited states is not confirmed experimentally. Here, we have taken $X=186$ keV \cite{Audi} for the experimental excitation energy of this state as shown in Fig. \ref{fig4}. In our calculation, the $\beta$-decaying isomer $(11/2^-)$ is proposed at 211 keV, which shows good agreement with the experimental level (at $X=186$ keV). The energies of other high-spin states are also close to the experimental energies of corresponding states except $23/2^-_1$. This state lies slightly lower than the experimental level. The  $11/2^-_1$, $15/2^-_1$, $19/2^-_1$, $23/2^-_1$, and $27/2^-_1$ states are formed by the coupling of the unpaired neutron in $\nu(h_{11/2})$ and $\pi (g_{9/2}^{-2})_{0^+,2^+,4^+,6^+,8^+}$.

{\bf $^{127}$Cd}: Our calculation reproduces the g.s. $(3/2^+)$ and the order of other experimental energy levels as in $^{125}$Cd. The locations of excited energy levels are also not confirmed by experiment in $^{127}$Cd. We have taken $X=283.3$ keV in the experimental energy spectra \cite{NNDC1}. The long-lived isomeric state $11/2^-$ \cite{Yordanov,Hoteling} lies 119 keV lower with our calculation in comparison to the experimental data. This state is formed with the contribution of a valence neutron in the $\nu(h_{11/2})$ orbital and is dominated by seniority $v=1$ ($\approx$56.8\%). The location of the tentative experimental energy states $(15/2^-)$ and $(19/2^-)$ \cite{Hoteling} are comparable with the experimental data from our calculations. The dominating configurations of the above two states consist of an odd neutron in $\nu(h_{11/2})$ which couple to $\pi(g_{9/2}^{-2})_{2^+}$ and $\pi (g_{9/2}^{-2})_{4^+}$, respectively.

{\bf $^{129}$Cd}: The experimental $11/2^-$ ground state is reproduced by the shell-model calculation. The location of the $3/2^+$ state is experimentally unconfirmed. In the experimental energy spectra, we have taken $X=0$ without any assumption. The calculated $3/2^+$ state lies 244 keV above the g.s., which may be helpful to compare with upcoming experimental data. With our calculation, the other excited states show good agreement with the experimental data. 
In Ref. \cite{Taprogge}, an isomeric state $27/2^-_1$ is predicted by the shell-model calculations below the calculated $21/2^+_1$ state. Our calculation supports their result and proposes that the $27/2^-_1$ state ($\pi(g_{9/2}^{-2})\otimes\nu(h_{11/2}^{-1})$) is formed by coupling between two proton holes in $\pi(g_{9/2})$ and one neutron hole in $\nu(h_{11/2})$ having seniority $v=3$.

\subsection{Deformation of the $11/2^-_1$ band}

Here, we focus on the $11/2^-_1$ states of Cd isotopes. First, we examine the band structure starting with the $11/2^-_1$ level. The comparison between the theoretical and experimental bands for $11/2^-$ is shown in Fig. \ref{band}. As shown later, the $23/2^-$--$19/2^-$--$15/2^-$--$11/2^-$ states are connected with strong $B(E2)$ transitions from the shell model. We can see that our calculated energies of the $11/2^-_1$ states show almost similar trends with the experiment. The energy of the $11/2^-_1$ state tends to decrease monotonically as the neutron number increases from the neutron-deficient region to the mid-shell. This decrease in energy is due to the fact that the neutron Fermi surface approaches the $h_{11/2}$ orbital in a similar way to the lowering of the $9/2^+_1$ levels in neutron-rich Cr, Fe, and Ni isotopes \cite{Togashi_Cr}.

In Fig. \ref{be2_comparison}, the calculated $B(E2; 15/2^-_1 \to 11/2^-_1)$ and $B(E2; 19/2^-_1 \to 15/2^-_1)$ values are depicted, together with the $B(E2; 2^+_1 \to 0^+_1)$ values for even-$A$ isotopes for comparison. One immediately finds that these two $B(E2)$ values in the odd-$A$ Cd isotopes exhibit a similar behavior to the $B(E2; 2^+_1 \to 0^+_1)$ values. The similarity between the $11/2^-_1$ band in odd-$A$ isotopes and the ground band in even-$A$ isotopes is also probed from the energy levels shown in Fig. \ref{band_comparision}. The excitation energies of the $15/2^-_1$, $19/2^-_1$, and $23/2^-_1$ states are taken with respect to the $11/2^-_1$ states are close to those of the $2^+_1$, $4^+_1$, and $6^+_1$ states of the neighboring nuclei. On the basis of these similarities, we propose that the $11/2^-_1$ band is dominated by the state that an unpaired neutron (a particle for lighter isotopes or a hole for heavier isotopes) in $h_{11/2}$ is coupled to the ground band of even-$A$ Cd isotopes. This situation is also similar to the $9/2^+_1$ bands in neutron-rich Cr, Fe, and Ni isotopes \cite{Togashi_Cr}.


\begin{figure}
	\includegraphics[width=0.2385\textwidth]{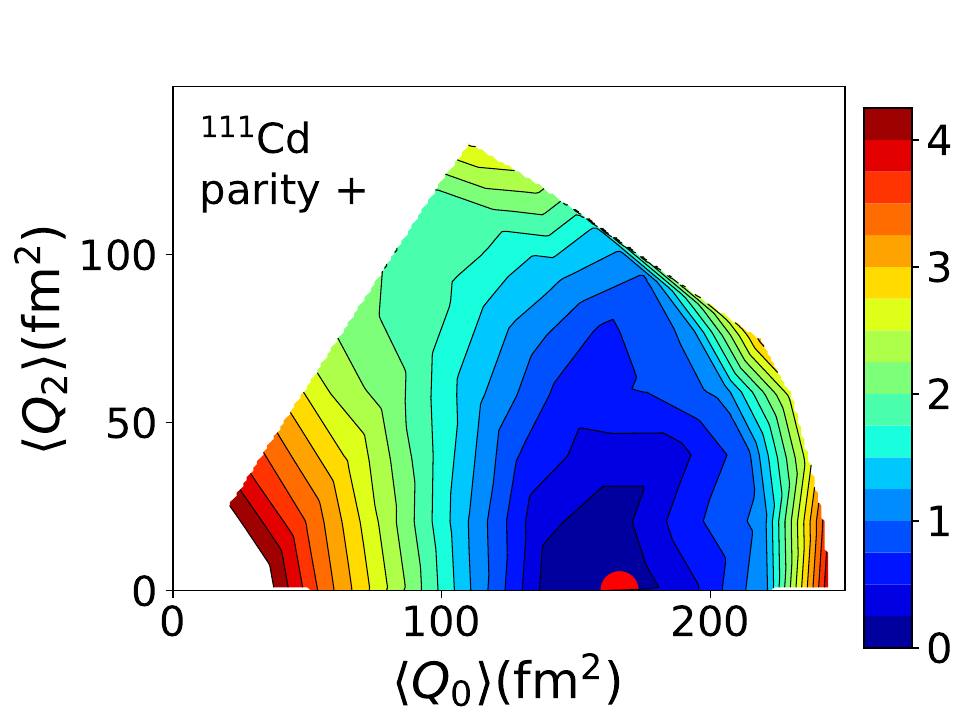}
	\includegraphics[width=0.2385\textwidth]{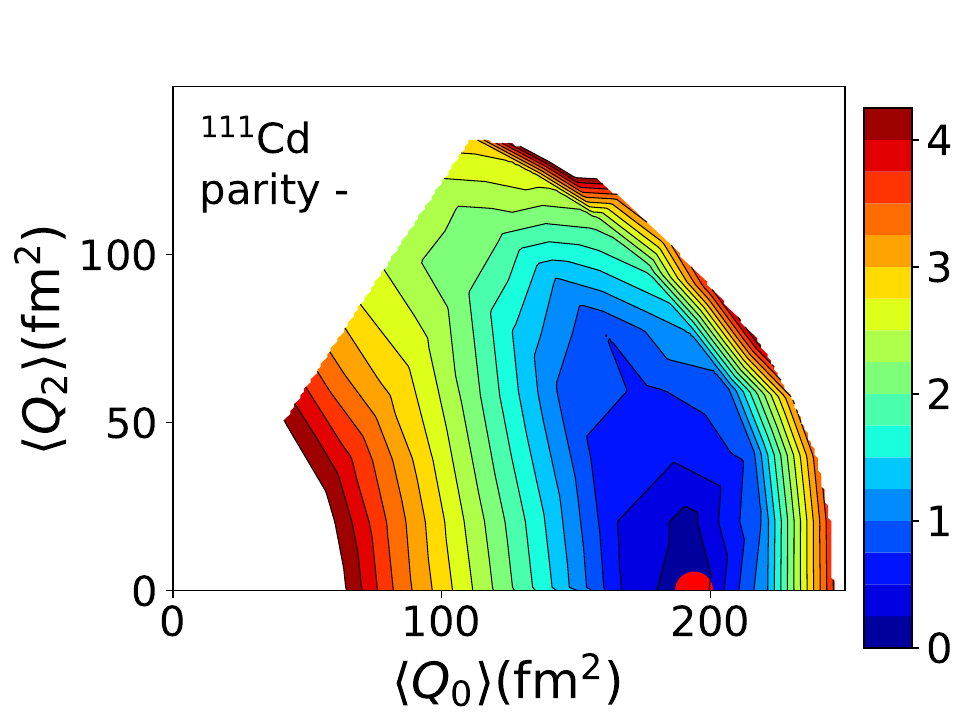}
    \includegraphics[width=0.2385\textwidth]{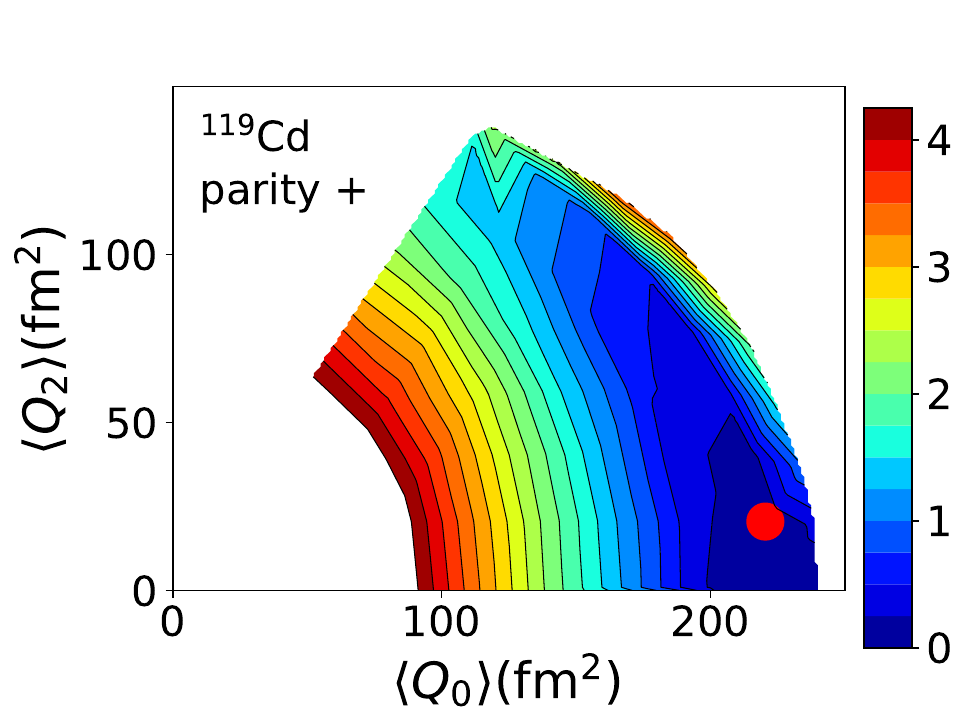}
    \includegraphics[width=0.2385\textwidth]{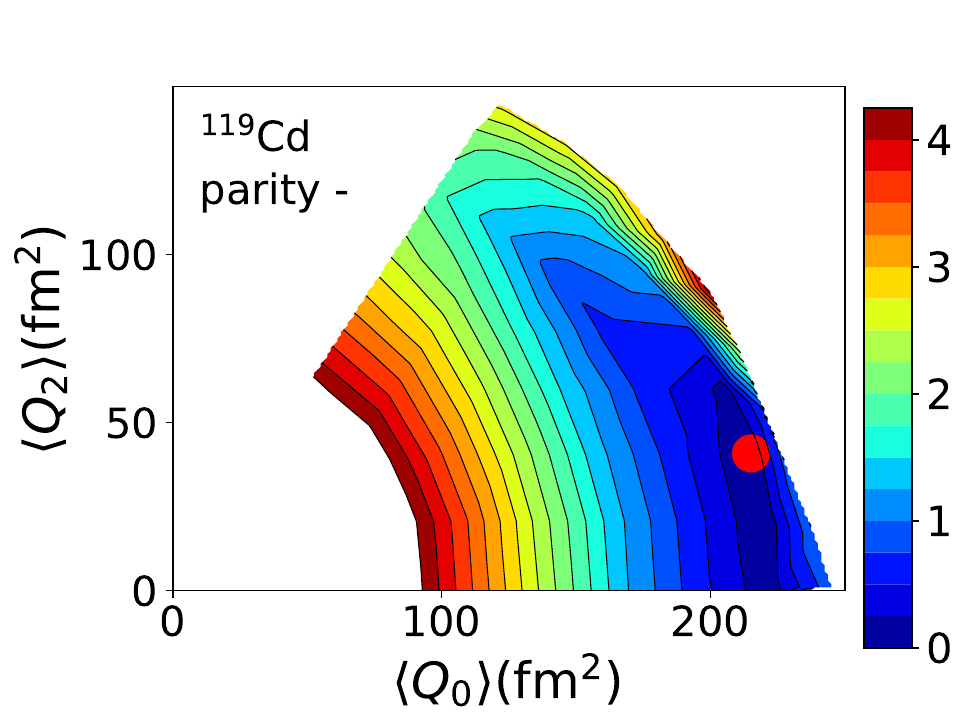}
    \includegraphics[width=0.2385\textwidth]{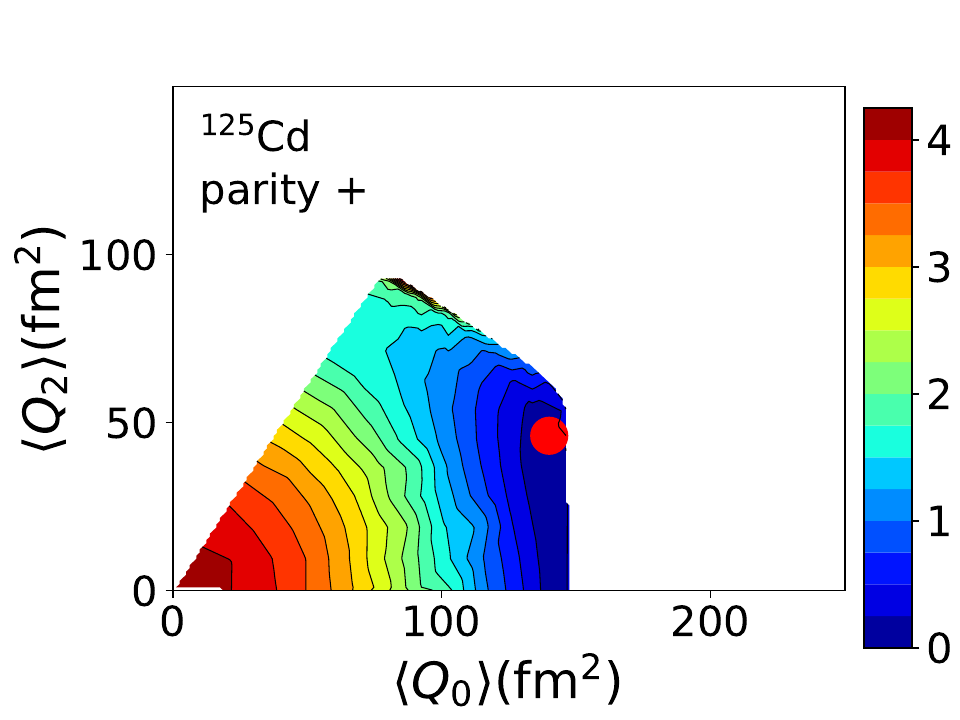}
    \includegraphics[width=0.2385\textwidth]{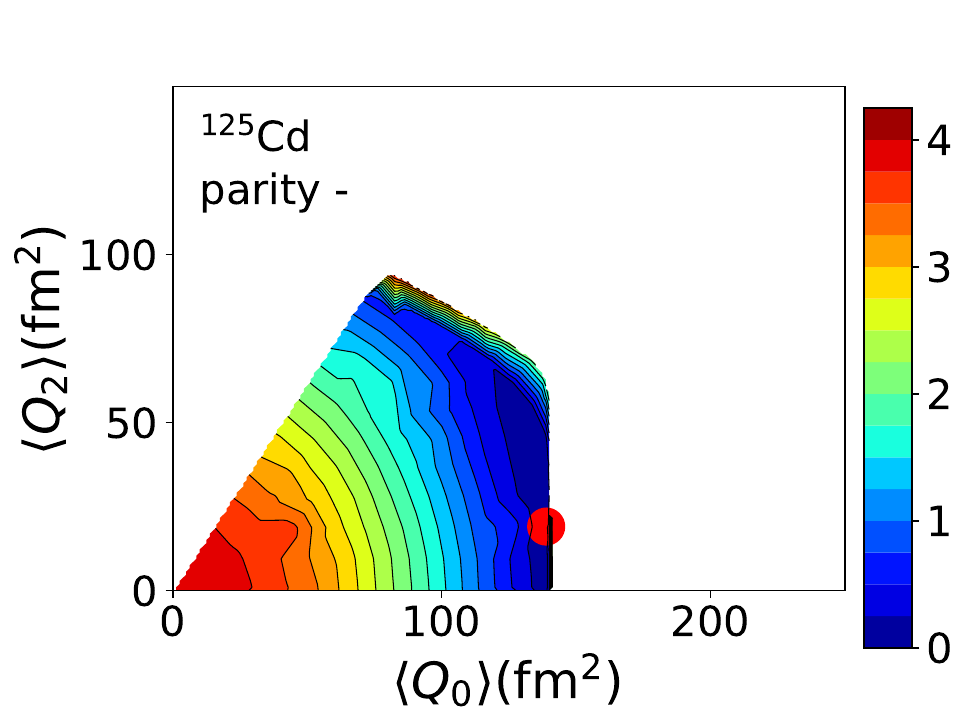}
    \includegraphics[width=0.2385\textwidth]{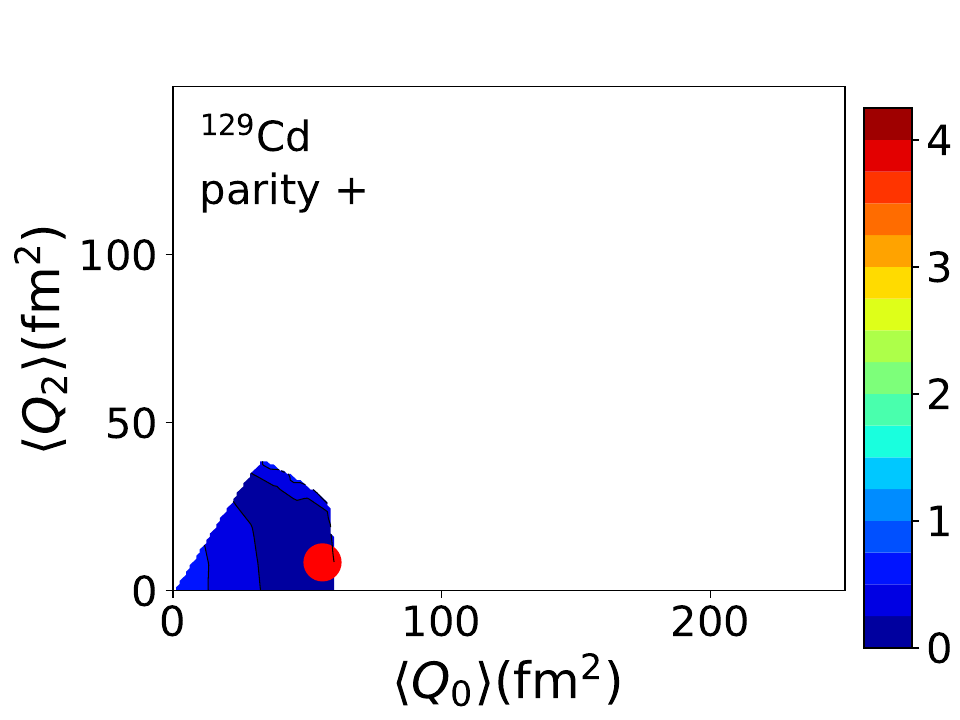}
    \includegraphics[width=0.2385\textwidth]{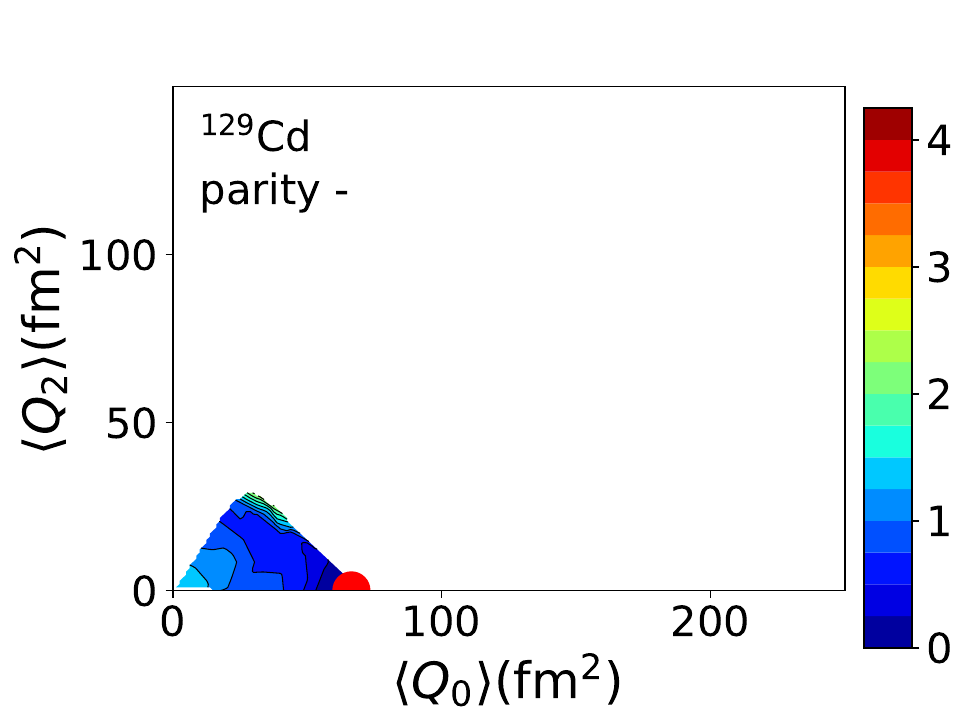}
	\caption{\label{fig15} Energy-surface plots for $^{111,119,125,129}$Cd isotopes. The red circles denote the energy minimum.}
\end{figure}

To probe nuclear shapes, the total energy surface is useful. Here we carry out the $Q$-constrained Hartree-Fock calculation with variation after parity projection that was employed in Ref. \cite{Togashi_Cr}. The model space and the effective interaction used for these calculations are the same as those of the shell-model calculations. The $11/2^-_1$ level is not the ground state, but is the lowest negative-parity state in the cases of interest. Hence the total energy surfaces with parity projected onto $+1$ and $-1$ well represent the optimum shapes of the ground state and the $11/2^-_1$ state, respectively. 

The results for $^{111,119,125,129}$Cd are depicted in Fig. \ref{fig15}. The total energy surfaces show that prolate shapes are favored both for the positive- and negative-parity states throughout the Cd isotope chain.
For $^{119,125,129}$Cd, the total energy becomes the lowest near the maximum deformation that the model space allows. This sometimes occurs when one employs the minimum model space to represent deformation. The deformation of the minimum energy peaks at around $N=70$, which is in accordance with the trend of the $B(E2)$ values presented in Fig. \ref{be2_comparison}. In addition, $\gamma$ instability develops with the neutron number.
Here, $\gamma$ is given by $\gamma=\arctan({\frac{\sqrt{2}\langle Q_2 \rangle}{\langle Q_0 \rangle}})$. 

While the positive-parity and negative-parity energy surfaces are rather similar, one can notice some differences in the optimum deformations. For $^{119,125}$Cd, those minima have different $\gamma$ deformations. But since the energy surface is rather shallow in the $\gamma$ direction in those cases, one should further investigate how this difference affects fully correlated wave functions. For $^{111}$Cd, on the other hand, the optimum $\langle Q_0 \rangle$ value for the negative-parity state is clearly larger than that for the positive-parity state. This result is consistent with the isomer shift observed recently \cite{Yordanov_isomer}.

\begin{figure}
	\includegraphics[width=85mm]{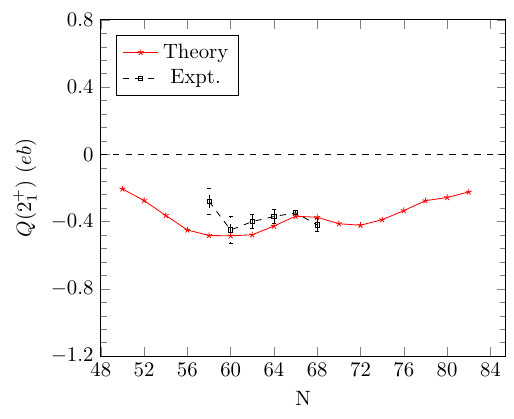}
	\caption{\label{even_q} Comparison between calculated and experimental \cite{NDS} Q($2^+_1$) for $^{98-130}$Cd isotopes.}
\end{figure}

For even-even nuclei, it is widely accepted that a negative $Q$ moment in the $2^+_1$ state points to a prolate shape, because the ground band is dominated by $K=0$. For Cd isotopes, the $Q(2^+_1)$ values are measured for $^{106}$Cd to $^{116}$Cd as plotted in Fig. \ref{even_q}. All the measured $Q$ moments are negative, which are well reproduced with our shell-model calculations. The absolute value of the $Q$ moment does not peak at around $N=70$ unlike the $B(E2)$ value. This may be due to triaxiality that develops for larger $N$ isotopes, because mixing with $K=2$ states leads to a positive shift in the $Q(2^+_1)$ moment. Anyway, the dominance of prolate shapes in Cd isotopes is thus supported by the experimental data. Since it is rather unlikely that adding a neutron drastically changes nuclear shape under the condition that the prolate shape is stable along the isotope chain, it is reasonable to claim the dominance of prolate shapes in the $11/2^-_1$ states.

\begin{table*}
	\centering
	\caption{Electric quadrupole and magnetic moments of odd-mass Cd isotopes. In our calculations we have taken ($e_p$, $e_n$)$=$(1.6, 0.8)$e$, $g_s^p=3.91$, and $g_s^n=-2.678$. In the fourth and seventh columns, the experimental data of quadrupole and magnetic moments, respectively, in $^{103-129}$Cd are taken from Ref. \cite{NDS}. For $^{101}$Cd, the experimental electromagnetic moments of $5/2^+_1$ state is taken from Ref. \cite{Yordanov1}. The fifth column shows the experimental quadrupole moments from Ref. \cite{Yordanov}.}
	\begin{ruledtabular}
		\begin{tabular}{lcccccccccccc}
			
			&          & \multicolumn{3}{c}{$Q$ ($e$b)}    & \multicolumn{2}{c}{$\mu$ ($\mu_N$)} \\ 
			\cline{3-5}
			\cline{6-7}
			
			~$A$  & $J^{\pi}$ & Theory &  Expt.  & Expt. & Theory & Expt.	\\ 
			\hline

			~99  & $5/2^+_{1}$  & -0.225 & - & -  & -1.132 & - \\ 
			
			101  & $5/2^+_{1}$  & -0.158 & -0.177(2) & -  & -0.785 & -0.8983(2) \\ 
			
			103  & $5/2^+_{1}$  & -0.096 & -0.007(3) & -  & -0.475 & -0.8486(4) \\ 
			
			105  & $5/2^+_{1}$  & -0.113 & 0.377(15) & -   & 0.632 & -0.7382(4) \\ 
			& $11/2^-_{1}$      & -0.753 &   -     & -   & -0.996  &   -  \\ 
			& $21/2^+_{1}$      & 0.919 & 0.88(10)  & -   & 9.153 & 9.17(6) \\ 
			
			107  & $5/2^+_{1}$  & 0.599 & 0.60(2)  & 0.601(3) &  0.219  & -0.6141(3) \\ 
			& $11/2^-_{1}$      & -0.782 & -0.81(10)  & -  & -0.990 & -1.041(11) \\ 
			& $21/2^+_{1}$      & 1.079 & 0.91(11)  & -   & 9.279 & 9.10(10) \\ 
			
			109  & $5/2^+_{1}$  & 0.620 & 0.60(3)  & 0.604(1) & -0.371 & -0.8266(4) \\ 
			& $11/2^-_{1}$      & -0.781 & -0.92(9) & -   & -1.023 & -1.096(2)\\ 
			
			111  & $1/2^+_{1}$  & - & - & -  & -0.294  & -0.5940(3) \\ 
			& $5/2^+_{1}$       & 0.621 & 0.64(3) & - & -0.166 & -0.766(3) \\ 
			& $3/2^+_{1}$       & -0.339 & - & -  & 0.914 & 0.9(6) \\ 
			& $11/2^-_{1}$      & -0.726 & -0.75(3) & -0.747(4) & -1.057 & -1.1036(5) \\ 

			113  & $1/2^+_{1}$  & - & - & - & -0.488 & -0.6213(3) \\ 
			& $11/2^-_{1}$      & -0.635 & -0.61(3) & -0.612(3) & -1.055 & -1.0883(3) \\ 
			& $3/2^+_{1}$       & -0.340 & - &  -   & 0.658 & -0.4(8) \\ 
			
			115  & $1/2^+_{1}$  & - & - & - & -0.689 & -0.6483(2) \\ 
			& $11/2^-_{1}$      & -0.479 & -0.48(2) & -0.476(5) & -1.040 & -1.0394(6) \\ 
			
			117  & $1/2^+_{1}$  & - & - & - & -0.682 & -0.7425(4)	 \\ 
			& $11/2^-_{1}$      & -0.306 & -0.320(13) & -0.320(6) & -1.011 & -0.9961(5) \\ 
			
			119  & $1/2^+_{1}$  & - & - & - & -0.627 & -0.9188(4)	\\ 
			& $11/2^-_{1}$      & -0.113 & -0.135(6) & -0.135(3) & -0.969 & -0.9628(5) \\ 
			
			121  & $3/2^+_{1}$  & -0.357 & - & -0.274(7)  & 0.469 & 0.6260(7) \\ 
			& $11/2^-_{1}$      & 0.058 & 0.009(6) & 0.009(6) & -0.942 & -1.0085(6) \\ 
			
			123  & $3/2^+_{1}$  & 0.257 & - & 0.042(5)   &  0.872  & 0.7885(7) \\ 
			& $11/2^-_{1}$      & 0.264 & 0.135(7) & 0.135(4) & -0.904 & -1.0000(6) \\ 
			
			125  & $3/2^+_{1}$  & 0.290 & 0.209(10) & 0.209(4) & 0.911 & 0.8591(7) \\ 
			& $11/2^-_{1}$      & 0.344 & 0.269(13) & 0.269(7)  & -0.912 & -0.9333(4) \\ 
			
			127  & $3/2^+_{1}$  & 0.246 & 0.239(11) & 0.239(5)  & 0.927 & 0.8771(8) \\ 
			& $11/2^-_{1}$      & 0.501 & 0.34(2) & 0.342(10)   & -1.001 & -0.8690(5) \\ 
			
			129  & $3/2^+_{1}$  & 0.160 & 0.132(9) & 0.132(7)   & 0.903 & 0.8469(9) \\ 
			& $11/2^-_{1}$      & 0.538 & 0.57(3) & 0.570(13)   & -0.908 & -0.7052(6) \\ 

		\end{tabular}
	\end{ruledtabular}
	\label{Moments}
\end{table*}

\subsection{Electromagnetic moments}

\begin{figure*}
 	\includegraphics[width=79mm]{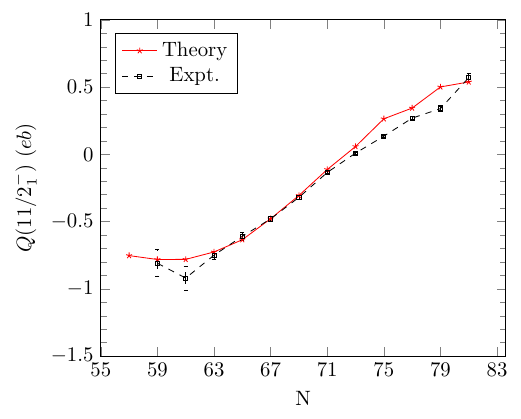}
	\includegraphics[width=79mm]{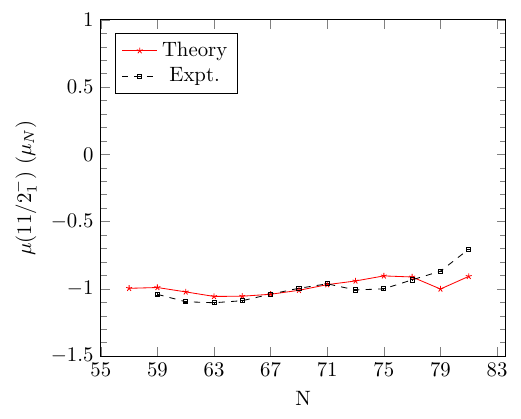}
	\caption{\label{quadrupole} Comparison between calculated and experimental \cite{NDS} electric quadrupole (left) and magnetic dipole (right) moments of the $11/2^-_1$ state for odd-mass $^{105-129}$Cd isotopes.}
\end{figure*}

\begin{figure}
 \includegraphics[width=93mm]{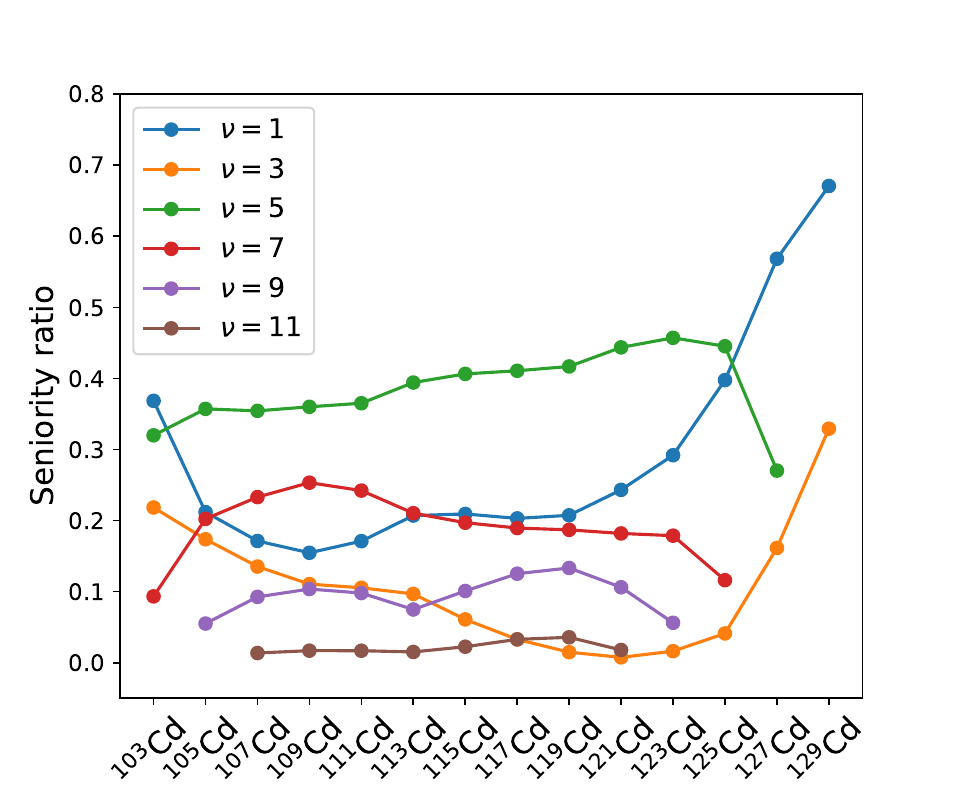}
	\caption{\label{Seniority} Seniority of the $11/2^-_1$ states of Cd isotopes.}
\end{figure}

We move on to electric quadrupole and magnetic moments in the odd-mass Cd isotopic chain reported in Table \ref{Moments}. The calculated moments are in reasonable agreement with the data on the whole. Here we focus on the systematics of electric quadrupole $(Q)$ and magnetic moment $(\mu)$ of the $11/2^-_1$ state, whose comparison between theory and experiment is presented in Fig. \ref{quadrupole}.

The quadrupole moment is frequently used to probe nuclear deformation. But its implication is not necessarily straightforward except for the ground bands of even-even nuclei, because the moment is measured in the laboratory frame, while deformation is defined in the intrinsic frame. For Cd isotopes, systematic data are available. As presented in Fig. \ref{quadrupole}, the $Q(11/2^-_1)$ value changes linearly with the neutron number. 
In terms of the single-$j$ shell model, the reduced 
matrix element of the $Q$ operator 
between a state with nucleon number $n$ and seniority $v$ in the orbital $j$ is described as 
\begin{equation}
    \langle j^n v  J|| Q || j^n v J '\rangle =\frac{2j+1-2n}{2j+1-2v}\langle j^{v} v J || Q || j^{v} v J' \rangle, 
    \label{eq:qn}
\end{equation}
with $J$ and $J'$ being the angular momenta. When the seniority is fixed, the quadrupole moment is proportional to $2j+1-2n$, thus leading to the linear behavior with $n$. Hence this linearity is often regarded as the success of the extreme shell model \cite{Yordanov}, which is useful for spherical nuclei. 

Our large-scale shell-model calculations also show reasonable agreement with the $Q(11/2^-_1)$ experimental trend, as shown in Fig. \ref{quadrupole}. Since the linearity based on Eq. (\ref{eq:qn}) is valid for states with a good seniority number, it is worth examining to what extent seniority is conserved in the large-scale shell-model results. 
The distribution of seniority numbers in the $11/2^-_1$ states is depicted in Fig. \ref{Seniority}. 
We can see that the lowest seniority $v=1$ is dominant only near $N=82$, 
whereas the seniority numbers are strongly mixed in the $11/2^-_1$ states for the other states; the largest fraction is $v=5$ in $^{105-125}$Cd.
The strong mixing in seniority is a natural consequence of the development of deformation in the $11/2^-_1$ states as discussed in the last subsection. 

\begin{figure}
	\includegraphics[width=79mm]{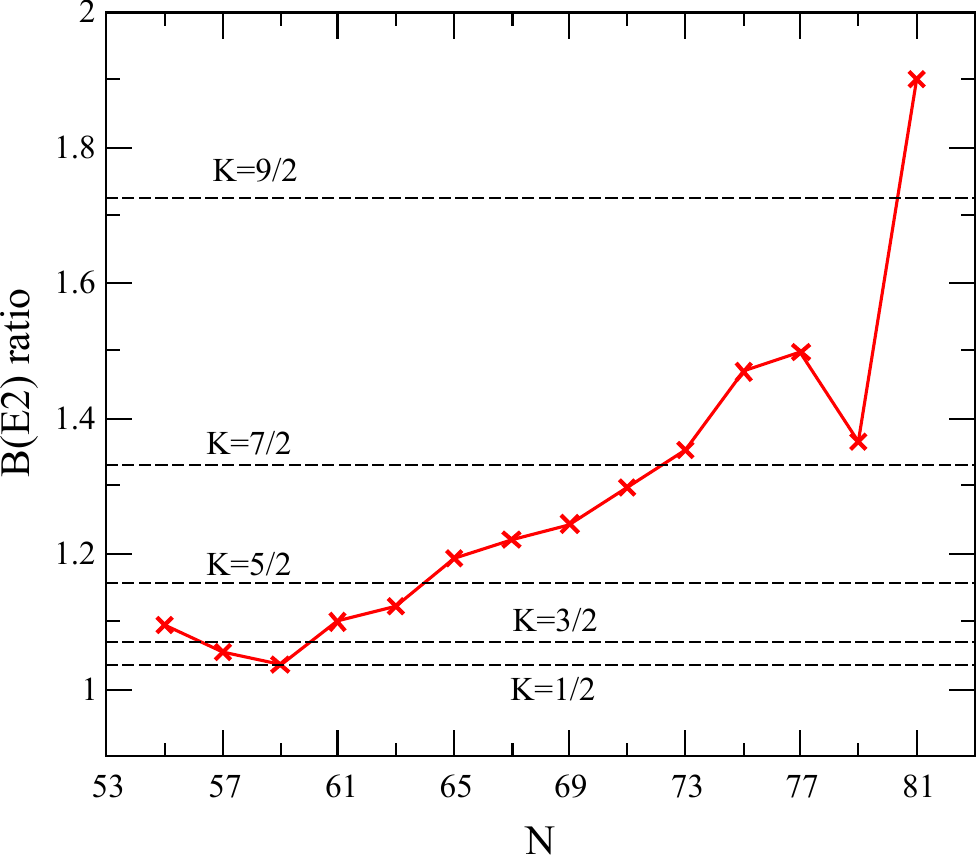}
	\caption{\label{be2ratio} The $B(E2)$ ratios $R$ [see Eq. (\ref{eq:ratio})] taken from the shell-model calculations (crosses) and those of fixed $K$ (dashed lines).}
\end{figure}

It is interesting that the $Q$ moments are similarly described with the different pictures, i.e., the extreme shell model that is based on spherical nuclei and the large-scale shell-model calculation that results in moderate prolate deformation. Here we show how the dominance of prolate shapes in Cd isotopes cause the observed $Q(11/2^-_1)$ values that change sign with the neutron number. If a nucleus is prolate deformed, the Nilsson orbitals originating from $h_{11/2}$ follow the order of $\Omega=1/2, 3/2, \ldots, 11/2$ from the lowest. The resulting $K$ number of a nucleus also follows this order with increasing neutron number because of the Pauli principle. When $K$ is a good quantum number, the spectroscopic (observed) $Q$ moment is related to the intrinsic $Q$ moment, $Q_0$, as 
\begin{equation}
    Q(J) = \frac{3K^2-J(J+1)}{(J+1)(2J+3)}Q_0 .
    \label{eq:qs}
\end{equation}
Thus when one increases $K$ from the lowest $K=1/2$ to the highest $K=J$ with $Q_0>0$ kept unchanged, the $Q(J)$ value changes sign from negative to positive.

Although $K$ numbers are not obtained with the shell-model calculation, the calculated $B(E2)$ values provide information on the changing $K$ numbers with the neutron number. If the $11/2^-$, $15/2^-$, and $19/2^-$ states form a rotational band with a fixed $K$ value, 
\begin{equation}
    R \equiv \frac{B(E2; 19/2^- \to 15/2^-)}{B(E2; 15/2^- \to 11/2^-)} 
    = \frac{(19/2\, K\, 2\, 0 | 15/2\, K)^2}{(15/2\, K\, 2\, 0 | 11/2\, K)^2} 
    \label{eq:ratio},
\end{equation}
which increases from 1.04 for $K=1/2$ to 3.04 for $K=11/2$ as a function of $K$. Hence the ratio of the calculated $R$ is a good measure for examining the dominant $K$. The shell-model values of $R$ are plotted in Fig. \ref{be2ratio}. Except for a few cases, the calculated $R$ value monotonically increases with the neutron number, in accordance with the increasing $K$ number.  

The $Q(11/2^-_1)$ value vanishes at around $N=73$ experimentally, whereas this happens for $K\approx 3.45$ from Eq. (\ref{eq:qs}). Such an effective $K$ number is realized at $N=71$-$73$ in terms of the analysis of the $B(E2)$ values presented in Fig. \ref{be2ratio}. This agreement confirms the validity of the simple argument based on the $K$ number. 

Another interesting systematics of the calculated $B(E2)$ values shown in Fig. \ref{be2_comparison} is that the $B(E2; 15/2^-_1 \to 11/2^-_1)$ value becomes smaller than the $B(E2; 2^+_1 \to 0^+_1)$ values of the neighboring nuclei for large neutron numbers. This is also understood by comparing the Clebsch-Gordan coefficients, $(15/2\, K\, 2\, 0 | 11/2\, K)$ and $(2\, 0\, 2\,0 | 0\, 0)$, the latter of which exceeds the former for $K=7/2$ and $K=9/2$.

As for the $Q$ moments of other states, Table \ref{Moments} shows that those of the $5/2^+_1$ state increase in $^{101-111}$Cd (except for $^{105}$Cd) with the increase of neutron number both experimentally and theoretically. The trend of $Q(5/2^+_1)$ is not necessarily linear, but it increases with the increasing neutron number.
The quadrupole moments of the $5/2_1^+$ states show very small variations for $^{107}$Cd and $^{109}$Cd. There are many similarities in the structure of both nuclei, which may cause the almost constant values of $Q(5/2^+_1)$ in both isotopes.

In the right panel of Fig. \ref{quadrupole}, the magnetic moment $\mu(11/2^-_1)$ is plotted from $N=57$ to $N=81$ for comparison between our calculated results and the experimental data. Our calculation reproduces the experimental trend of $\mu(11/2^-_1)$, suggesting the dominance of an unpaired neutron in the $h_{11/2}$ orbital. Similarly, the magnetic moments of the $3/2^+_1$ states are rather stable along the isotope chain (except for $^{113}$Cd with a large experimental uncertainty) due to the dominance of an unpaired neutron in $d_{3/2}$. The calculated magnetic moments of the $21/2^+$ isomer in $^{105,107}$Cd are also in good agreement with the experimental data.

\subsection{Isomeric states in neutron-rich In isotopes}

\begin{figure*}
	\includegraphics[width=173mm]{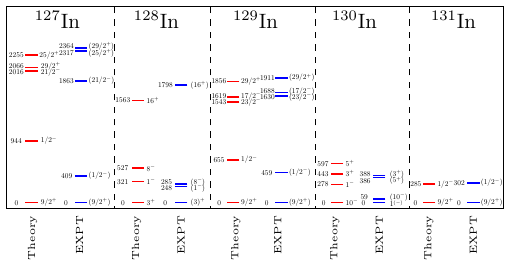}
	\caption{\label{In_spectra} Comparison between calculated and experimental \cite{NNDC} energy levels for $^{127-131}$In isotopes.}
\end{figure*}

In this subsection, we discuss the isomeric states in $^{127-131}$In isotopes. The shell-model calculations for the In isotopes were performed using the same interaction as in the previous subsections. 
The experimental data for energy spectra are taken from Ref. \cite{Nesterenko}, in which shell-model study of $^{128,130}$In with the jj45pna interaction is also reported.
Figure \ref{In_spectra} shows the excitation spectra for isomeric states in $^{127-131}$In isotopes. The dominating configurations and seniority of isomeric states are reported in Table \ref{isomer}. 
The shell-model electromagnetic moments of the ground and the isomeric states are compared with the experimental data in Table \ref{Moments_In}.

\begin{table}
	\caption{Dominant configurations of isomeric states in neutron-rich In isotopes and their seniority percentages of the shell-model wave functions.}
	\begin{ruledtabular}
		\begin{tabular}{ccccc}
			Isotope	& $J^{\pi}$ & Seniority (\%)   & Dominant configuration \\
			\hline
			&  &      &   \\
			$^{127}$In	& $1/2^-_1$  & $v=1$ (80.8)   & $\pi(p_{1/2}^{-1})$ \\
			& $21/2^-_1$    & $v=5$ (69.7)   & $\pi(g_{9/2}^{-1})\otimes\nu(d_{3/2}^{-1}h_{11/2}^{-1})$ \\
			& $29/2^+_1$    & $v=3$ (78.7)   & $\pi(g_{9/2}^{-1})\otimes\nu(h_{11/2}^{-2})$\\
			&  &      &   \\            
			\hline
			
			&  &      &   \\
			$^{128}$In & $1^-_1$    & $v=2$ (71.0)  & $\pi(g_{9/2}^{-1})\otimes\nu(h_{11/2}^{-1})$ \\
			& $8^-_1$     & $v=4$ (58.5)  & $\pi(g_{9/2}^{-1})\otimes\nu(h_{11/2}^{-1})$ \\
			& $16^+_1$    & $v=4$ (100)  & $\pi(g_{9/2}^{-1})\otimes\nu(d_{3/2}^{-1}h_{11/2}^{-2})$\\
			&  &      &   \\            
			\hline   
			
			&  &      &   \\
			$^{129}$In	& $1/2^-_1$     & $v=1$ (87.6)  & $\pi(p_{1/2}^{-1})$\\
			& $23/2^-_1$    & $v=3$ (100) & $\pi(g_{9/2}^{-1})\otimes\nu(d_{3/2}^{-1}h_{11/2}^{-1})$ \\
			& $17/2^-_1$    & $v=3$ (100) & $\pi(g_{9/2}^{-1})\otimes\nu(d_{3/2}^{-1}h_{11/2}^{-1})$ \\
			& $29/2^+_1$    & $v=3$ (100)  & $\pi(g_{9/2}^{-1})\otimes\nu(h_{11/2}^{-2})$ \\	
			&  &      &    \\            
			\hline   
			
			&  &      &   \\
			$^{130}$In	& $10^-_1$  & $v=2$ (100)  & $\pi(g_{9/2}^{-1})\otimes\nu(h_{11/2}^{-1})$  \\
			& $5^+_1$    & $v=2$ (100) & $\pi(p_{1/2}^{-1})\otimes\nu(h_{11/2}^{-1})$  \\
			& $3^+_1$    & $v=2$ (100) & $\pi(g_{9/2}^{-1})\otimes\nu(d_{3/2}^{-1})$\\
			&  &      &   \\            
			\hline   
			
			&  &      &   \\
			$^{131}$In	& $1/2^-_1$  & $v=1$ (100)   & $\pi(p_{1/2}^{-1})$\\

		\end{tabular}
	\end{ruledtabular}
	\label{isomer}
\end{table}

\begin{table*}
	\centering
	\caption{Electromagnetic moments of ground and isomeric states in neutron-rich In isotopes. 
  The effective charges are taken as $(e_p, e_n)=(1.6, 0.8)e$ for the quadrupole moments, and the spin $g$ factor  is quenched by a factor 0.7 for the magnetic moments. 
 The experimental data are taken from Ref. \cite{Vernon,Vernon1}.}
	\begin{ruledtabular}
		\begin{tabular}{lcccrccccc}
			&             & \multicolumn{2}{c}{$Q$ ($e$b)} & \multicolumn{2}{c}{$\mu$ ($\mu_N$)} \\ 
			\cline{3-4}
			\cline{5-6}
			
			Isotope  & $J^{\pi}$  & Theory  & Expt. \cite{Vernon,Vernon1}

			& Theory & Expt. \cite{Vernon,Vernon1}
			\\ 
			\hline
			
			$^{127}$In  & $9/2^+_{1}$  & 0.602 & 0.588(29) & 5.399 & 5.5321(14) \\ 
			& $1/2^-_{1}$  & - & - & -0.323 & -0.4355(24) \\ 
			& $21/2^-_{1}$  & 0.778 & 0.806(22) & 5.303 & 5.3398(38) \\ 
			& $29/2^+_{1}$  & 0.751 & - & 3.432 & - \\ 
			
			$^{128}$In  & $3^+_{1}$  & 0.431 & 0.416(22) & 3.916 & 4.0883(19) \\ 
			& $1^-_{1}$  & -0.019 & - & -2.710 & - \\ 
			& $8^-_{1}$  & 0.431 & 0.462(21) & 3.658 & 4.6549(33) \\ 
			& $16^+_{1}$  & 0.764 & - & 4.607 & - \\ 
			
			$^{129}$In  & $9/2^+_{1}$  & 0.487 & 0.487(13) & 5.517 & 5.5961(23) \\ 
			& $1/2^-_{1}$  & - & - & -0.200 & -0.38709(58) \\ 
			& $23/2^-_{1}$  & 0.672 & 0.598(32) & 5.666 & 5.9349(39) \\ 
			& $17/2^-_{1}$  & 0.606 & - & 3.463 & - \\ 
			& $29/2^+_{1}$  & 0.639 & - & 3.521 & - \\ 
			
			$^{130}$In  & $1^-_{1}$  & 0.069 & 0.0599(96) & -3.568 & -3.4542(34) \\ 
			& $10^-_{1}$  & 0.545 & 0.578(26) & 4.616 & 5.0138(44) \\ 
			& $5^+_{1}$  & 0.309 & 0.372(15) & 2.224 & 2.8686(18) \\ 
			& $3^+_{1}$  & 0.316 & - & 4.441 & - \\ 
			
			$^{131}$In  & $9/2^+_{1}$  & 0.336 & 0.31(1) & 5.955 & 6.312(14) \\ 
			& $1/2^-_{1}$  & - & - & 0.015 & -0.051(3) \\ 

		\end{tabular}
	\end{ruledtabular}
	\label{Moments_In}
	
\end{table*}

Our results successfully reproduce the experimentally assigned $9/2^+$ ground states in the odd-mass $^{127-131}$In isotopes, which are dominated by the $\pi(g_{9/2})^{-1}$ configuration. In the case of $^{131}$In, the excitation energy of the isomeric state $(1/2^-)$ is comparable with the experimental data from our calculation, while the $1/2^-$ excitation energies of $^{127}$In and $^{129}$In are rather overestimated. These $1/2^-_1$ states are mainly dominated by the $\pi(p_{1/2}^{-1})$ configuration.

The intruder $\nu(h_{11/2})$ orbital lies just below the $N=82$ shell closure, and thus plays an essential role in forming high spin isomeric states in neutron-rich odd-mass indium isotopes, as shown in Table \ref{isomer}. We discuss these isomeric states in terms of seniority quantum number \cite{Racah,Flowers}. Seniority ($v$) is the number of particles not in pairs coupled to the angular momentum $J=0$. The shell-model configurations of the $17/2^-$, $23/2^-$, and $29/2^+$ isomers of $^{129}$In are dominated by the seniority $v=3$ and are mainly formed by $[(\pi g_{9/2}^{-1}\nu h_{11/2}^{-1})_{10}\otimes \nu d_{3/2}^{-1}]_{17/2^{-},23/2^{-}}$ and $[(\pi g_{9/2}^{-1})\otimes(\nu h_{11/2}^{-2})_{10}]_{29/2^{+}}$. These orbits are also involved in forming the $29/2^+$ state of $^{127}$In. Only $v=3$ configurations are allowed in the $23/2^-_1$, $17/2^-_1$, $29/2^+_1$ states of $^{129}$In, as shown in Table \ref{isomer}. In $^{127}$In and $^{129}$In, the isomeric $29/2^+$ states decays via $E2$ and $E3$ transitions, respectively. In $^{129}$In, shell-model predicted $B(E3;29/2^+\rightarrow23/2^-)$ = 57.6 $e^2$fm$^6$ shows a reasonable agreement with the experimental value, 68.2(99) $e^2$fm$^6$ \cite{NNDC_NUDAT}. However, for $^{127}$In the calculated $B(E2;29/2^+\rightarrow25/2^+)$ transition probability, 103.8 $e^2$fm$^4$, deviates from the experimental value, 11.4(27) $e^2$fm$^4$ \cite{NNDC_NUDAT}.

Our shell-model study indicates that the $1^-$ and $3^+$ states of $^{128}$In are dominated by the $v=2$ configurations, namely $[\pi(g_{9/2}^{-1})\otimes\nu(h_{11/2}^{-1})]$ and  $[\pi(g_{9/2}^{-1})\otimes\nu(d_{3/2}^{-1})]$, respectively. 
Concerning the excitation energy of the $\beta-$decay $8^-$ isomer of $^{128}$In, our result shows better agreement with the experimental one than the previous shell-model result with the jj45pna interaction \cite{Nesterenko}.
While the major component of the $8^-$ state is $\pi(g_{9/2}^{-1})\otimes \nu(h_{11/2}^{-1})$, the main configuration of the new isomeric $16^+$ state is proposed as $[(\pi g_{9/2}^{-1} \otimes \nu d_{3/2}^{-1})_{6^+}\otimes (h_{11/2}^{-2})_{10^+}]_{16^+}$. 
In the present work, it is not possible to predict the properties of two high spin isomers $21/2^+$ and $17/2^+$ in $^{131}$In with the present model space because the neutron excitation across the $N=82$ gap needs to be included.

We also compare our calculated magnetic ($\mu$) and quadrupole ($Q$) moments of the ground and isomeric states in $^{127-131}$In with the available experimental data \cite{Vernon,Vernon1,Flynn}. The indium isotopes are unique cases to study the effect of coupling of a single proton hole with different neutron particles for both observables ($\mu$ and $Q$). In Table \ref{Moments_In}, there is very small variation in $\mu(9/2^+)$ value from $^{127}$In to $^{129}$In, while accountable change occurs at $^{131}$In. The shell-model value of the magnetic moment of the $^{131}$In $9/2^+$ ground state is  $\mu = $ 5.955 $\mu_N$, which is given by the Schmidt value of the $\pi(g_{9/2})$ orbit with the spin $g$ factor quenched by 0.7. The experimental $Q(9/2^+)$ value decreases gradually from $^{127}$In to $^{131}$In. The shell-model results for $Q(9/2^+)$ also follow the same trend. The $Q(9/2^+)$ value of $^{131}$In is smaller than $^{127,129}$In, which means that the $^{131}$In is less deformed, since it is an $N=82$ submagic nucleus. The shell-model $\mu$ and $Q$ values are also comparable for the other isomeric states in $^{127-131}$In. The experimental data of quadrupole and magnetic moments for several states are not yet available, and our shell-model predictions might be useful to compare with upcoming experimental data.

\section{Summary and Conclusions} \label{section4}

We have performed large-scale shell-model calculations for odd-mass Cd isotopes for $N=51-81$. 
Except for the ground-state spin-parities of some nuclei, 
the shell-model results show reasonable agreement 
with the experimental data including the energy spectra and electromagnetic properties. 
On the basis of the good agreement, we have investigated what causes the quadrupole moments in the $11/2^-$ states that linearly change with increasing neutron number. In contrast to the conventional understanding in terms of the single-$j$ shell model with the seniority scheme, we propose that the $11/2^-_1$ states have prolate deformation throughout the Cd isotope chain. This result is taken from the $Q$-constrained Hartree-Fock calculations, and is also supported by the observed $Q$ moments in even-$A$ Cd isotopes. 
The $Q(11/2^-_1)$ values change as observed due to the changing $K$ numbers of neutrons that fill from the lowest $\Omega$ Nilsson orbitals. This picture also well accounts for the fact that $B(E2; 2^+_1 \to 0^+_1)$ peaks at around $N=70$, which may be difficult to obtain in the seniority scheme or its variants. 
We have also analyzed different isomeric states in neutron-rich $^{127-131}$In isotopes in terms of shell-model configurations and seniority quantum number $(v)$. We have discussed the importance of the $\nu(h_{11/2})$ orbital in forming the high-spin isomers in indium isotopes. We have also studied the behavior of magnetic and quadrupole moments of ground and isomeric states in indium isotopes and given the prediction of $\mu$ and $Q$ values where experimental data are not available.\\

\section*{\uppercase{Acknowledgements}}

We acknowledge N. Smirnova and M. Honma for providing us their effective shell-model interactions. We acknowledge financial support from SERB (India), Grant No. CRG/2022/005167. D.P. acknowledge financial support from MHRD, the Government of India. We would like to thank the National Supercomputing Mission (NSM) for providing computing resources of ``PARAM Ganga'' at the Indian Institute of Technology Roorkee, implemented by C-DAC and supported by MeitY and DST, Government of India. N.S. and Y.U. acknowledge the support of ``Program for promoting researches on the supercomputer Fugaku,'' MEXT, Japan (Grant No. JPMXP1020230411), and the support of JSPS KANENHI Grant No. 20K03981. N.S. acknowledges the MCRP program of the Center for Computational Sciences, University of Tsukuba (NUCLSM).

\end{document}